\documentclass[aps,pre,reprint,showpacs,floatfix,superscriptaddress,nofootinbib]{revtex4-1}
\usepackage{graphicx}
\usepackage{subfigure}
\usepackage{url}
\usepackage{hyperref}
\usepackage{inconsolata}
\usepackage{amsmath}	%for align environment and other things
\allowdisplaybreaks  	%so aligned equations can break across pages
\usepackage[capitalize]{cleveref}
\crefname{section}{Sec.}{Secs.}
\crefname{appendix}{App.}{Apps.}
\usepackage{braket}
\usepackage{color}

\usepackage{bm} % bold math
\DeclareMathOperator{\Tr}{Tr}
\usepackage[utf8]{inputenc}
\usepackage{siunitx,booktabs}
\usepackage{multirow}
\usepackage{amssymb}

\begin{document}
\bibliographystyle{apsrev4-2}
%\bibliographystyle{aipnum4-1} % this one uses the URLs to make bibliography references clickable links (Awesome)
% Use the \preprint command to place your local institutional report
% number in the upper righthand corner of the title page in preprint mode.
% Multiple \preprint commands are allowed.
% Use the 'preprintnumbers' class option to override journal defaults
% to display numbers if necessary
%\preprint{}

%Title of paper
\title{Bose-Einstein condensation of deconfined spinons in two dimensions}

% repeat the \author .. \affiliation  etc. as needed
% \email, \thanks, \homepage, \altaffiliation all apply to the current
% author. Explanatory text should go in the []'s, actual e-mail
% address or url should go in the {}'s for \email and \homepage.
% Please use the appropriate macro foreach each type of information

% \affiliation command applies to all authors since the last
% \affiliation command. The \affiliation command should follow the
% other information
% \affiliation can be followed by \email, \homepage, \thanks as well.
\author{Adam Iaizzi}
\email{iaizzi@bu.edu}
\homepage{www.iaizzi.me}
\affiliation{Department of Physics, National Taiwan University, No. 1, Section 4, Roosevelt Road, Taipei 10607, Taiwan}
\affiliation{Department of Physics, Boston University, 590 Commonwealth Avenue, Boston, MA 02215, USA}

\author{Harley D. Scammell}
\email{harleyscammell@gmail.com}
\affiliation{School of Physics, The University of New South Wales, Sydney, New South Wales 2052, Australia}
\affiliation{Department of Physics, Harvard University, Cambridge MA 02138, USA}

\author{Oleg P. Sushkov}
\email{sushkov@unsw.edu.au}
\affiliation{School of Physics, The University of New South Wales, Sydney, New South Wales 2052, Australia}

\author{Anders W. Sandvik}
\email{sandvik@bu.edu}
\affiliation{Department of Physics, Boston University, 590 Commonwealth Avenue, Boston, MA 02215, USA}
\affiliation{Beijing National Laboratory for Condensed Matter Physics and Institute of Physics, Chinese Academy of Sciences, Beijing 100190, China}

%Collaboration name if desired (requires use of superscriptaddress
%option in \documentclass). \noaffiliation is required (may also be
%used with the \author command).
%\collaboration can be followed by \email, \homepage, \thanks as well.
%\collaboration{}
%\noaffiliation

\date{\today}
\begin{abstract}
The transition between the N\'{e}el antiferromagnet and the valence-bond solid state in two dimensions has become a paradigmatic example of deconfined 
quantum criticality, a non-Landau transition characterized by fractionalized excitations (spinons). We consider an extension of this scenario whereby the deconfined spinons are subject to a magnetic field. 
The primary purpose is to identify the exotic scenario of a Bose-Einstein condensate of spinons. 
We employ quantum Monte Carlo simulations of the \mbox{$J$-$Q$} model with a magnetic field and perform a quantum field theoretic analysis of the magnetic field and temperature dependence of thermodynamic quantities. 
The combined analysis provides evidence for Bose-Einstein condensation of spinons and also demonstrates an extended temperature regime in which the system is best described as a gas of spinons interacting with an emergent gauge field.
\end{abstract}

%\maketitle must follow title, authors, abstract, \pacs, and \keywords
\maketitle

%\tableofcontents
%\vfill
%\pagebreak

%=================================================
\section{Introduction}

Symmetry-breaking phase transitions are normally described by the Landau-Ginzburg paradigm in which the critical point is governed by the order parameter of the ordered phase. 
A notable conclusion of Landau theory is that phase transitions between states breaking unrelated symmetries should be first order. 
In the past two decades large-scale quantum Monte Carlo (QMC) results \cite{sandvik2007,melko2008,lou2009,jin2013,chen2013,pujari2013,harada2013,shao2016} have uncovered evidence of a new type of critical point that violates this rule: the apparently continuous transition between the O(3) N\'{e}el antiferromagnet (AFM) and the Z$_4$ valence-bond solid (VBS) in 2D quantum magnets \cite{sandvik2007}. 
This transition is believed to be an example of \textit{deconfined quantum criticality} (DQC), a type of non-Landau transition where the critical point is described not by the order parameter of either ordered phase, but by emergent fractionalized excitations that appear only near the DQC point (in this case spinons, $S=\frac{1}{2}$ bosons) \cite{senthil2004,senthil}. 
The critical system can be described as a U(1) spin liquid \cite{ma2018}.

In this paper, we extend the study of deconfined spinons to include an external magnetic field. 
The field extends the critical point to a line separating the VBS and a field-induced Bose-Einstein condensate (BEC) \cite{scammell2015}. 
As we will show, the field forces a finite density of magnetic excitations into the ground state and drives them to form a BEC, which changes to an interacting gas at higher temperatures. 
The low-temperature behavior of spinons is different from magnons (the conventional $S=1$ excitation of an AFM \cite{fisher1989,affleck1991}). 
We predict how they will differ using a quantum field theory analysis of spinons, including a crucial dynamical gauge field that was neglected in previous work \cite{scammell2015}. 
We then compare the theory to large-scale QMC simulations, demonstrating an excellent match to the spinon theory and the failure of the magnon theory. 
In particular, the effects of the emergent gauge field remain large at temperatures well above the BEC transition temperature. 

\textit{Outline:} 
The background and context for this work are discussed along with our methods in \cref{s:back}. 
Section \ref{s:becBound} contains our estimation of the BEC phase boundary. 
We then describe our field theoretical approach and our evidence for a spinon BEC and spinon gas in \cref{s:qft}. 
We have provided more detailed derivations in the appendices: 
in \cref{supp:fittingDetails} we extract the parameters for our field theory;
in \cref{supp:renormDetails} we derive the perturbative loop corrections to the Green's functions and partition function;  
in \cref{supp:totEn} we calculate the energy predictions from the spinon theory; In
\cref{supp:symmetry} we describe the symmetry breaking and obtain dispersions of all modes; 
finally, in \cref{supp:magnon} we describe the magnon theory. 

%=================================================
\section{Background \label{s:back}}

%\textit{Background.}---
The VBS is a nonmagnetic phase characterized by a long-range ordered arrangement of local singlets breaking Z$_4$ lattice symmetry \cite{majumdar1,majumdar2,haldane1982,haldane1988,read1989,sachdev2008}. 
VBS physics is traditionally studied in frustrated systems \cite{dagotto1989,schultz1996}, but these suffer from QMC sign problems.
Fortunately, many aspects of frustrated systems can be mimicked with other types of competing interactions. 
Here we use the 2D square-lattice \mbox{$J$-$Q$} model \cite{sandvik2007}, a sign-problem-free Hamiltonian formed by augmenting a Heisenberg term of strength $J$ with a four-spin interaction of the form $-QP_{i,j}P_{k,l}$, where $P_{i,j}$ is a singlet projection operator \mbox{$P_{ij}\equiv \frac{1}{4} - \mathbf{S}_i \cdot \mathbf{S}_j$} 
with $S=\frac{1}{2}$: 
\begin{equation}
H = -J \sum \limits_{\braket{i,j}} P_{i,j} - Q \sum \limits_{\braket{i,j,k,l}} P_{i,j} P_{k,l} + h \sum \limits_i S^z_i. \label{eq:JQh}
\end{equation}
Here $\braket{i,j}$ represents nearest neighbors and $\braket{i,j,k,l}$ correspond to $2\times2$ plaquettes with indices arranged both as $\begin{smallmatrix} k&l\\i&j\end{smallmatrix} $ and $\begin{smallmatrix} j&l\\i&k\end{smallmatrix}$ (preserving all lattice symmetries).
We fix $Q=1$ and refer to the dimensionless coupling ratio $j\equiv J/Q$. 
For zero field, the $Q$ term drives a transition from N\'{e}el AFM to VBS at $j_c \approx 0.045$ \cite{suwa2016,shao2016}. 

There is still some debate as to whether this transition is truly continuous or merely weakly first order (perhaps connecting to an inaccessible non-unitary 
critical point \cite{wang2017}). 
It is nonetheless clear that the spinons are deconfined up to a very large length scale, such that many unconventional aspects of the DQC theory appear. 
Their hallmarks can be seen in the thermodynamics at zero field 
\cite{sandvik2011} as well as in the dynamical properties \cite{ma2018}; therefore, deconfined spinons (as opposed to magnons) are the appropriate degrees of freedom to describe this transition.  
Our approach using an external field has several advantages: the field allows for direct control of the density of magnetic excitations and thus allows for the formation of a BEC. 
Furthermore, the field alters the dispersion of the low-lying modes, thereby producing much clearer signatures of deconfinement in the leading-order temperature dependence than in the zero-field case \cite{sandvik2011}. 

%=================================================
%\section{Methods \label{s:methods}}

\textit{Methods}---Our numerical results were generated using the stochastic series expansion QMC method \cite{sandvik2011computational} with directed loop updates \cite{sandvik_dl} and $\beta$-doubling \cite{sandvik2002} based on a method used in our previous work \cite{iaizzi2015,iaizzi2017,iaizzi2018metamag}. 
These techniques are described in a detailed manner for the specific model considered here in Ref.~\onlinecite{mythesis}. 

%=================================================
\section{BEC Phase Boundary \label{s:becBound}}

\begin{figure}
\includegraphics[width=87mm]{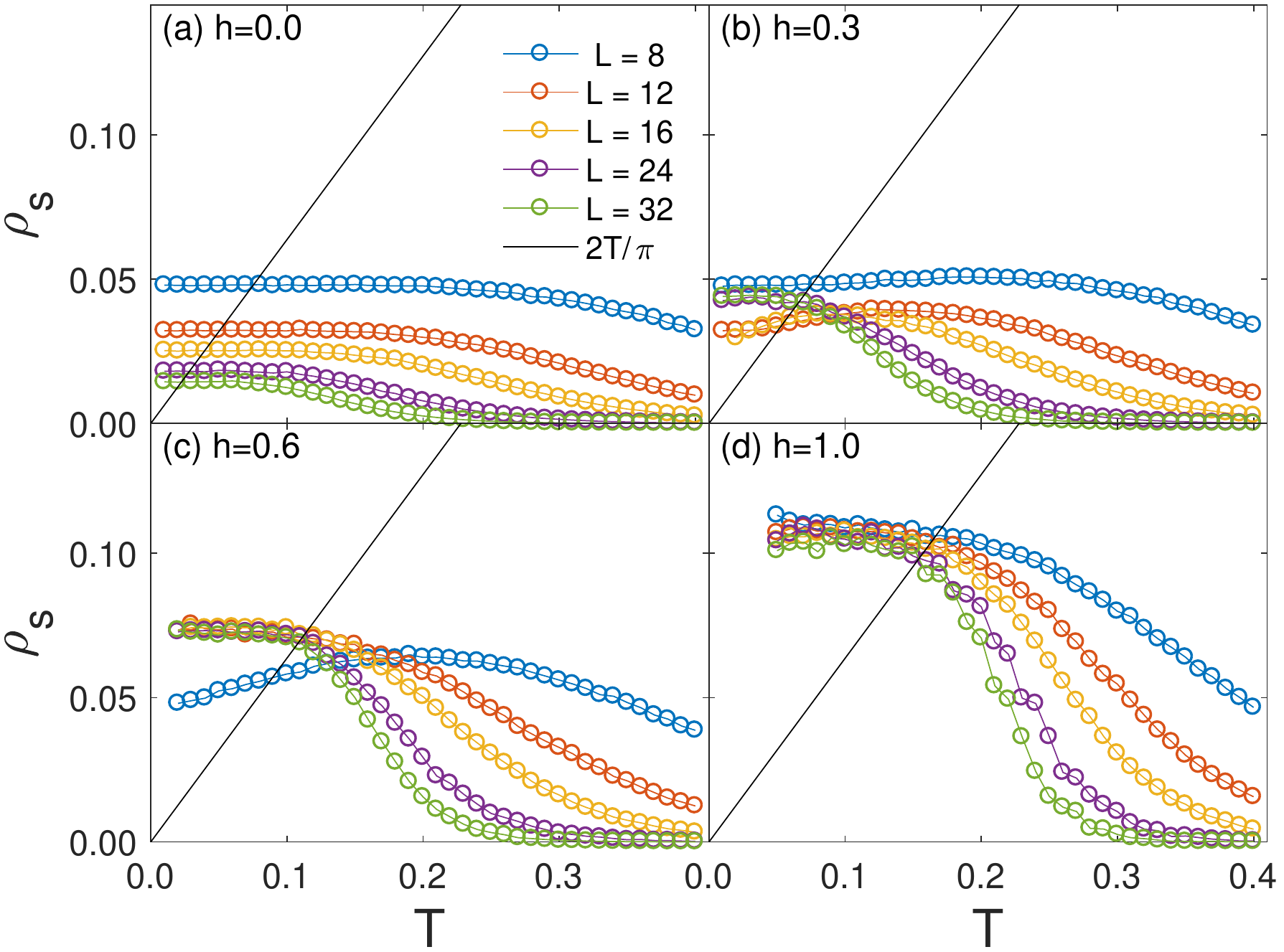}
\caption{Finite size scaling of the stiffness $\rho_s(T,h)$ from QMC simulations with \mbox{$j=j_c=0.045$} for (a) $h=0$, (b) $h=0.3$, (c) $h=0.6$ and (d) $h=1.0$. Error bars are omitted for clarity but are smaller than or equal to the markers. The black lines show the Nelson-Kosterlitz criterion, \cref{e:nelson}. Note the non-monotonic size dependence at low $T$ in (b) and (c). \label{f:rho}}
\end{figure}

\begin{figure}[t]
\includegraphics[width=70mm]{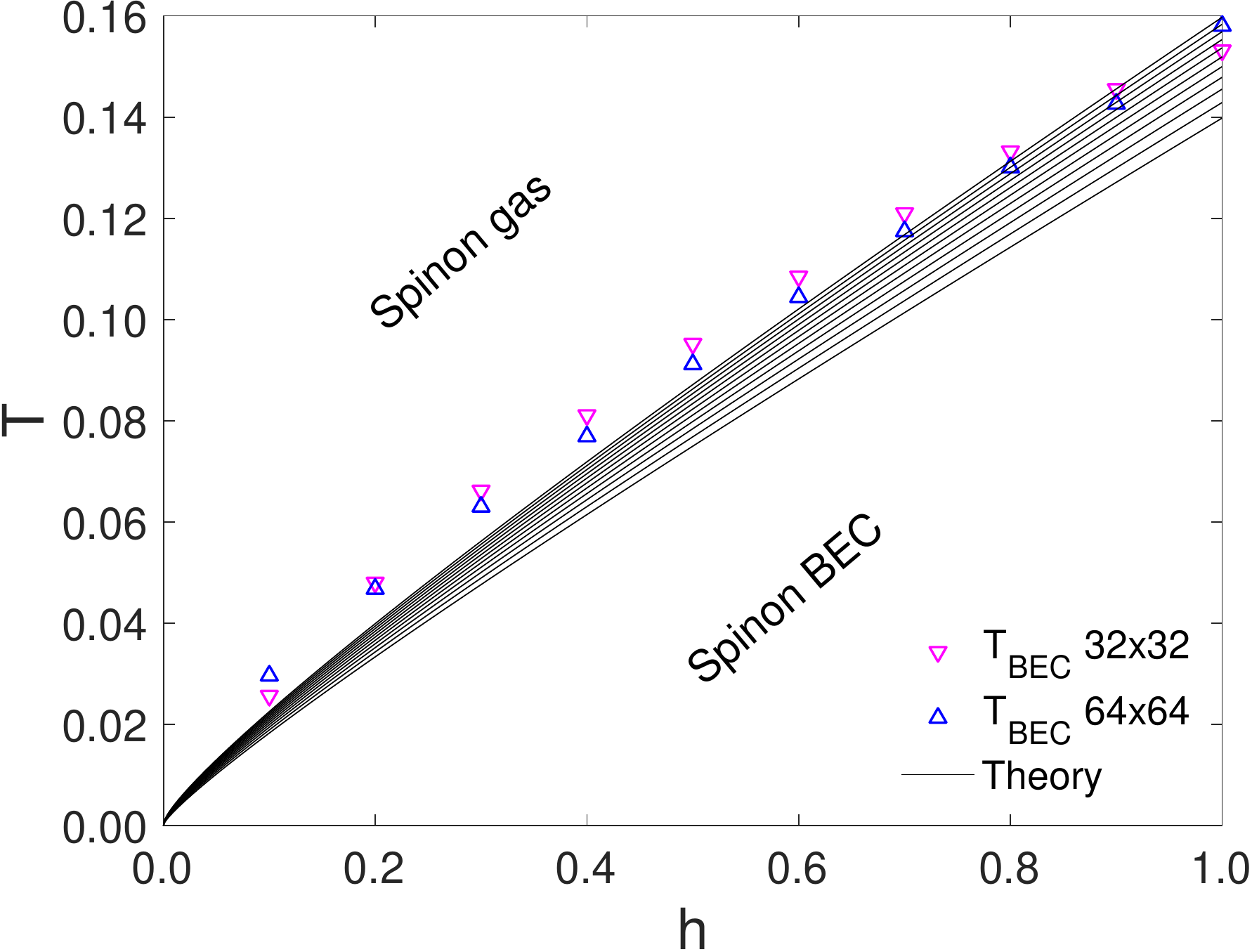}
\caption{Phase diagram of the $J$-$Q$ model in the $h$-$T$ plane with \mbox{$j=j_c=0.045$}. Triangles represent the QMC values of $T_{\rm BEC}(h)$ extracted from $\rho_s$. Fine lines represent phase boundaries based on the field theoretic solution to \mbox{$\Delta_T(h_c)-\mu h_c=\delta$}, as defined by \cref{gapdef}, with each curve based on a different $\delta\in\{10^{-6},10^{-4}\}$; the predicted transition temperature decreases as $\delta$ decreases (see \cref{s:gas}).  \label{f:pd2}}
\end{figure}

%\textit{BEC phase boundary.}---
The magnetic field forces a nonzero density of magnetic excitations into the ground state. 
At low temperature, these excitations form a BEC. 
Strictly, no long-range order is formed at $T>0$ as this  is prohibited by the Mermin-Wagner theorem, so this state may not meet the most stringent definition of a BEC.
However, the quasi-BEC state is still a `stiff' state as demonstrated in \cref{f:rho}. 
Above $T_{\rm BEC}(h)$, defined as per \cref{f:pd2}, the excitations have the character of a gas. 
An important aspect of our work is also to understand the nature of this interacting gas.

In terms of the spin lattice model [\cref{eq:JQh}] the transition between quasi-BEC and gas is analogous to the Berezinkii-Kosterlitz-Thouless (BKT) transition in the 2D classical XY model \cite{kosterlitz1972}, since the external field explicitly breaks the full SU(2) rotational symmetry of the spins to in-plane `XY' symmetry \cite{mythesis,landau1981,pires1994,cuccoli2003,cuccoli2004,baranova2016}. 
The $J$-$Q$ model under applied field is related to the anisotropic $J$-$Q$ model; hosting the same rotational symmetries, but lacking particle-hole symmetry. 
The $\rm XY\rightarrow Z_4$ transition in the anisotropic $J$-$Q$ model has also been shown to be direct and continuous (or possibly weakly first order), and therefore is also amenable 
to a spinon treatment \cite{qin2017,ma2018,ma2019}. 

We determine $T_{\rm BEC}(h)$ using the spin stiffness $\rho_s$, which measures the energy cost of a long-range twist about the $S^z$ axis \cite{sandvik2011computational,mythesis}. 
We show finite size scaling of $\rho_s(T,h)$ near the DQC point in \cref{f:rho} for \mbox{$h=0,0.3,0.6,1$}. 
For $h=0$, there is no BEC and $\rho_s$ vanishes as $L\rightarrow\infty$. 
For $h=0.3$, $\rho_s$ is finite as $L\rightarrow \infty$, reflecting the onset of a stiff phase, but the finite size scaling is nontrivial; as a function of $L$, $\rho_s(h\neq0)$ first \textit{decreases} and then \textit{increases} towards an asymptote. 
This behavior reflects the competition between the effects of finite size and finite temperature pushing the system towards the different phases near the multicritical point. 
For higher fields this non-monotonic behavior is less prevalent. 
For all $h\neq0$ the finite-size effects quickly become unimportant at larger sizes. 

We determined $T_{\rm BEC}(h)$ using the Nelson-Kosterlitz criterion, 
\begin{equation}
\rho_s (T_{\rm BEC}) = \frac{2 T_{\rm BEC}}{\pi}, \label{e:nelson}
\end{equation}
which governs the onset of a BKT transition \cite{nelson1977,hsieh2013}. 
We plot \cref{e:nelson} as a black line in \cref{f:rho}. 
To minimize error, we fit a polynomial to our QMC results $\rho_s(T,h)$ in the region around in crossing with \cref{e:nelson} and then solve the polynomial for $T_{\rm BEC}(h)$. 
The results of this procedure, $T_{\rm BEC}(h)$ for $L=32, \, 64$, are presented in \cref{f:pd2} (along with field theory estimates of the crossover temperature, described in \cref{s:gas}). 
Although the finite-size effects have not completely converged by $L=64$, they are sufficiently converged for our purposes here, which require only a rough estimate of the boundary between the BEC regime and the gas regime. 
From the presence of this stiff phase we conclude that there is a BEC of magnetic excitations (of some kind) for $T<T_{\rm BEC}(h)$. 
This approach cannot \textit{a priori} tell us whether the condensing excitations are spinons or conventional magnons. 
For that, we turn to a field theory description of spinons. 

%=================================================
\section{Results \label{s:qft}}

\begin{figure} 
\includegraphics[width=80mm]{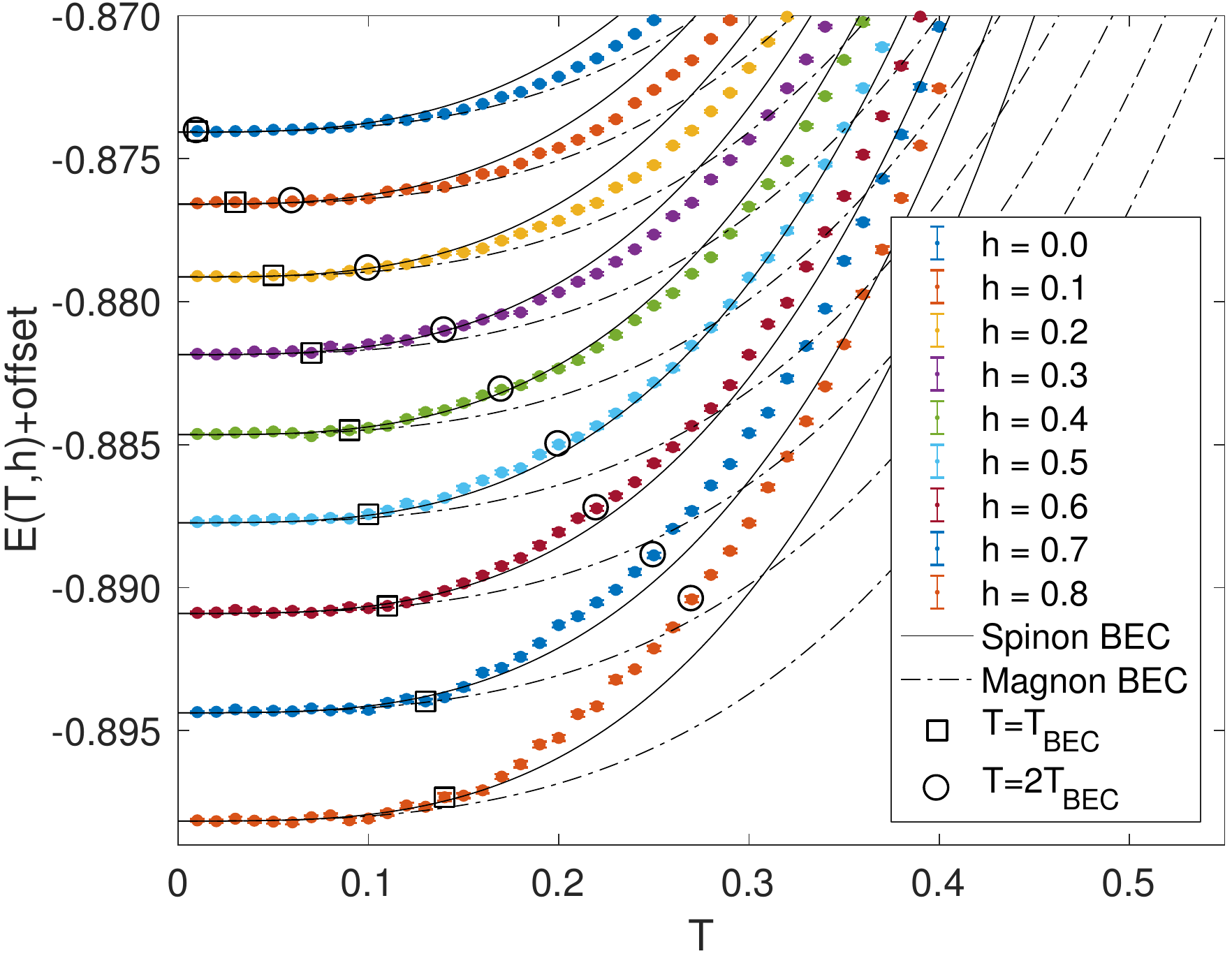}
\caption{BEC of spinons for $T<T_{\rm BEC}(h)$. Colored $\bullet$ are QMC results for $E(T,h)+$offset compared to the field theory predictions for a BEC of spinons (solid line) and a BEC of magnons (broken line). The points $E(T_{\rm BEC},h)$, $E(2T_{\rm BEC},h)$ are marked with $\square$, $\circ$ respectively. Theory lines are numerically exact; QMC results' error bars are smaller than the markers. \label{f:bec}}
\end{figure}

%\textit{Quantum field theory.}---
We adopt a bosonic field theory approach and work directly with deconfined spinon excitations in a $(2+1)d$ quantum field theory. 
In the Lagrangian, spinons $(z)$ are minimally coupled to a deconfined U(1) dynamical gauge field $(a_\nu)$ \cite{senthil,senthil2004} with an external magnetic field $(\vec{h})$ coupled to spin, but not charge: 
\begin{align}
\mathcal{L}=&\mathcal{L}[z]+\mathcal{L}[a_\nu]+\mathcal{L}[z,a_\nu], \label{eq:L} \\
\mathcal{L}[z]=&\{\partial_t z^\dag + i \mu z^\dag(\vec{\sigma}\cdot\vec{h})\}\{\partial_t z - i \mu(\vec{\sigma}\cdot\vec{h})z\}\nonumber\\*
&-c^2(\nabla z^\dag)(\nabla z) -\Delta_0^2z^\dag z - \frac{\alpha}{2}(z^\dag z)^2, \nonumber\\
\mathcal{L}[a_\nu]=&-\frac{1}{4}\cal{F}_{\mu\nu}\cal{F}^{\mu\nu}, \nonumber \\
\mathcal{L}[z,a_\nu]=&i e a^\nu (z^\dag\partial_\nu z - \partial_\nu z^\dag z) -e^2a_\nu a^\nu z^\dag z \nonumber \\* %prevents break after specific line
&+ \mu e(\vec{\sigma}\cdot\vec{h}z + z^\dag\vec{\sigma}\cdot\vec{h})a_\nu, \nonumber
\end{align}
where $\mu=1/2$ is spin of the spinon, $\Delta_0$ is the $T=0$ spinon mass, and $c$ is the spinon velocity (which also applies to the gauge field). 
In \mbox{$(2+1)d$}, the indices $\mu,\nu=0,1,2$, and the coupling constants have dimensions of energy: $\alpha=\tilde{\alpha}\Lambda$, $e^2=\tilde{e}^2\Lambda$ such that  
$\tilde\alpha, \tilde e$ are dimensionless couplings and $\Lambda$ is an infrared energy scale. 
This scale is $\Lambda=\Delta_0$ in the gas, and \mbox{$\Lambda=\mu h$} in the BEC. 

For the spinon velocity we use the previously extracted value ($c=2.42$, converted to our units from Ref. \onlinecite{suwa2016}). 
The remaining phenomenological field theory parameters are determined by fitting to our own QMC results (see \cref{supp:fittingDetails}). 
We determine dimensionless coupling ratios $\{\tilde\alpha, \tilde{e}\}=\{\frac{2}{3}\pi c^2 (0.32),0.75c\}$ by fitting to the QMC condensate energy (as a function of magnetic field)\footnote{We did this fitting for \mbox{$T=0.05$}, the lowest temperature studied with QMC} and magnetic susceptibility (as a function of temperature). 
Obtaining $\{\tilde\alpha, \tilde{e}, c\}$ fixes all free parameters. 

%=================================================
\subsection{Spinon BEC \label{s:bec} }

%\textit{Spinon BEC.}---
For $T < T_{\rm BEC}$ the Lagrangian [\cref{eq:L}] describes a BEC-like phase with order parameter, condensate energy, and magnetization given by 
\begin{equation}
\rho^2 \equiv z_0^\dag z_0 = \frac{\mu h}{\tilde{\alpha}}, \
{\cal E}_0=-\frac{\mu^3 h^3}{2\tilde{\alpha}}, \ 
\braket{m}=-\frac{\partial{\cal E}_0}{\partial h},  \label{eq:condensate} 
\end{equation}
respectively. 
Because we are at the DQC point, we explicitly set the spinon mass $\Delta =\Delta_0 = 0$. 
The spinon BEC has five modes: 
\begin{small}
\begin{align}
\omega^2_{1}&=3\mu^2 h^2+c^2k^2-\sqrt{(3\mu^2 h^2)^2+4\mu^2 h^2c^2k^2}, \notag \\
\omega^2_{2}&=3\mu^2 h^2+c^2k^2+\sqrt{(3\mu^2 h^2)^2+4\mu^2 h^2c^2k^2}, \notag \\
\omega^2_{3}&=2\mu^2 h^2 +e^2\rho^2+c^2 k^2-\sqrt{\left(e^2\rho^2-2 \mu^2 h^2\right)^2+4 \mu^2 h^2 c^2 k^2}, \notag \\
\omega^2_{4}&=2\mu^2 h^2 +e^2\rho^2+c^2 k^2+\sqrt{\left(e^2\rho^2-2 \mu^2 h^2 \right)^2+4 \mu^2 h^2 c^2 k^2}, \notag \\
\omega^2_{5}&=c^2k^2+2e^2\rho^2. \label{BECspec}
\end{align}% comments to eliminate indent after equation
\end{small}%
As a function of applied magnetic field, all five modes are continuous across the BEC transition, as expected for a second order transition.
There are two pure spinon modes unaltered by the gauge field: $\omega_{1}$, a gapless linear Goldstone mode and $\omega_2$, a gapped mode. 
$\omega_{3}$ and $\omega_{4}$ are hybrid spinon-gauge modes, with gaps given by
\begin{subequations}
\begin{align}
\Delta_3& =
\begin{cases}
  \sqrt{2}e\rho, & e^2\rho^2<2 \mu^2 h^2,\\    
2\mu h, & e^2\rho^2>2 \mu^2 h^2.
\end{cases}\\
\Delta_4& =
\begin{cases}
  2\mu h, & e^2\rho^2<2 \mu^2 h^2,\\    
\sqrt{2}e\rho, & e^2\rho^2>2 \mu^2 h^2. 
\end{cases}
\end{align}
\end{subequations}
There is always one gap of $2\mu h$ and a second of $\sqrt{2}e\rho$, but which mode ($\omega_3$ or $\omega_4$) has which gap depends on the relative magnitude of $e^2\rho^2$ and $2 \mu^2 h^2$ (this is a result of the anti-crossing behavior of the coupled modes). 
Finally, $\omega_{5}$ is a gauge mode which is gapped by the Anderson-Higgs mechanism. 

We have gone beyond our previous work \cite{scammell2015} by including coupling to a U(1) gauge field in our field theory. 
The previous study assumed this field to be unimportant because it is not necessary to describe the zero-field behavior \cite{sandvik2011}. 
In fact, the gauge field dramatically alters the modes, most importantly by destroying the gapless quadratic mode. 
The new hybridized spinon-gauge mode $(\omega_3)$ has a small gap (compare to $\omega_{1k}$ in Eq.~(8) of Ref.~\onlinecite{scammell2015}).
As we will show, this changes the leading-order temperature dependence of the thermodynamic energy from $E\propto T^2$ to $E\propto T^3$ as $T\to0$. 

We obtain the energy in the predicted modes from the partition function $E(h,T)=-T^2\partial_T\ln Z$. 
The dominant energy contributions in the spinon BEC come from the gapless Goldstone mode $\omega_1$ and the almost-gapless hybrid spinon-gauge mode $\omega_3$. 
Since neither of these modes are quadratic, there is no anomalous leading-order temperature dependence  $E\propto T^2$ as $T\to0$ \cite{scammell2015}.  

Deep in the BEC phase, both the spinon and magnon theories host a single (linear in $k$) Goldstone mode; details of the magnon theory are left for \cref{supp:magnon}. 
The linear modes of each theory have identical effective velocity, i.e.
\begin{align}
\omega_1=\omega_G\to \sqrt{\frac{\mu^2 h^2-\Delta^2}{3\mu^2 h^2-\Delta^2}}ck=\frac{c}{\sqrt{3}}k,
\end{align} 
as $k\to0$ (using $\Delta=0$ to obtain the RHS). 
This is $\omega_1$ of \cref{BECspec} for the spinon theory and $\omega_G$ of \cref{dispPI} for the magnon theory. 
This gapless linear mode dominates the energy $E(h,T)$, and its contribution is given by  
\begin{align}
E_\text{lin}(h,T)&\approx\frac{T^3}{c^2}\frac{3}{\pi}\zeta(3).
\end{align}
Hence the spinon and magnon theories approximately coincide deep in the BEC (i.e. as $T\ll h$), as seen in \cref{f:bec}. 
Importantly, however, there are significant differences between the spinon and magnon theories near the BEC phase transition temperature $T\sim T_\text{BEC}$. 
Here the linear mode becomes quadratic,
\begin{align}
\omega_1=\omega_G\to \sqrt{\frac{2(\mu^2 h^2)^2}{(3\mu^2 h^2-\Delta^2)^3}}c^2k^2=\frac{\sqrt{2}c^2}{3\mu h}k^2,
\end{align} 
as $k\to0$, and again using that $\Delta=0$ in the equality. 
This holds for both spinons $\omega_1$ and magnons $\omega_G$ [\cref{dispPI}]. 
The spinon theory hosts a second mode, $\omega_3$ of \cref{BECspec}, which becomes gapless and quadratic at $T=T_\text{BEC}$, 
\begin{align}
\omega_3\to \frac{c^2}{2\mu h}k^2,
\end{align} 
as $k\to0$. 
There is no analog of this quadratic mode in the magnon theory. 
This distinguishes the two theories, since the gapless quadratic modes are the dominant contributions to the energy at $T\sim T_{BEC}$. 
Evaluating the contribution from the gapless quadratic modes for the magnon and spinon theories, 
\begin{align}
E^\text{magnon}_\text{quad}(h,T)&\approx T^2\frac{\mu h}{c^2}\frac{\pi}{24}\frac{3}{\sqrt{2}},\\
E^\text{spinon}_\text{quad}(h,T)&\approx T^2\frac{\mu h}{c^2}\frac{\pi}{24}\left(\frac{3}{\sqrt{2}}+ 2\right),
\end{align}
we see that the contribution from these modes in the spinon theory is almost twice as large. 
This accounts for much of the differences observed in \cref{f:bec} in the vicinity of $T\sim T_\text{BEC}$.

In \cref{f:bec} we compare QMC results for $E(T,h)$ to the theory predictions for a BEC of spinons and a BEC of magnons. 
Each line is offset by $F(h)=-0.025h$ to prevent the curves from overlapping. 
All free parameters in the theory were fixed by fitting to other quantities, so no fitting has been performed in this figure other than to shift the theory curves to pass through the corresponding QMC data at \mbox{$T=0.05$}. 
We expect the BEC theory to describe the system from $T=0$ to around $T \approx T_{\rm BEC}(h)$, above which the BEC is no longer the mean field ground state. 
This is indeed what we observe. 

Unfortunately, the $E(T,h)$ predictions from the spinon and magnon BEC theories are very similar, which makes it difficult to draw a solid conclusion about the nature of the excitations from the behavior of the BEC alone.  
This is expected: deep within the BEC ($h\gg T \approx0$) the spinons are reconfining, therefore the magnon and spinon theories will coincide. 
The statistical energy is dominated by the gapless modes, especially at low temperature. 
Deep in the ordered phase, the spinon and magnon theories both have a single gapless (linear) Goldstone mode and, therefore, have the same leading temperature dependence. 
The differences between the spinon and magnon BEC predictions are in the \textit{sub}leading terms. 
Only near the transition, as the additional modes of the spinon theory begin to soften, can one expect significant deviation between the two theories. 
Indeed, in \cref{f:bec} we see that near $T_{\rm BEC}$ the spinon theory is a better match to the QMC data compared to the magnon theory. 
In the next section we discuss the gas regime $T\gtrapprox T_{\rm BEC}$, where the difference between the spinon and magnon theories is more dramatic. 

%=================================================
\subsection{Spinon Gas \label{s:gas} }

%\textit{Spinon Gas.}---
For $T\gtrsim  T_{\rm BEC}$ the magnetic excitations form a gas instead of a BEC. 
In this phase, the condensate order parameter $(\rho^2)$ and condensate energy $({\cal E}_0)$ vanish. 
There are five modes:  
\begin{subequations}%
\begin{align}%
\omega_{-}&=\sqrt{\Delta_T^2+c^2k^2} -{\mu h}   &&\times2, \label{eq:spinonGasModes1} \\
\omega_{+}&=\sqrt{\Delta_T^2+c^2k^2} +{\mu h}   &&\times2,\\
\omega_{\gamma}&=c k,
\end{align}\label{eq:spinonGasModes}\end{subequations}
two pairs of degenerate spinon modes $(\omega_{\pm})$, and the U(1) gauge mode $(\omega_{\gamma})$, which in this case does \textit{not} hybridize with the spinon modes. 
Unlike the BEC, the spinon gas modes have a $T$, $h$, and $k$-dependent {\it thermal mass}, $\Delta_T$, due to interactions with the gauge field and self-interactions. 
The gas appears when the thermal mass $\Delta_T>\mu h$, as is evident from the gapless point in \cref{eq:spinonGasModes1}. 

\begin{figure}[t] 
\includegraphics[width=80mm]{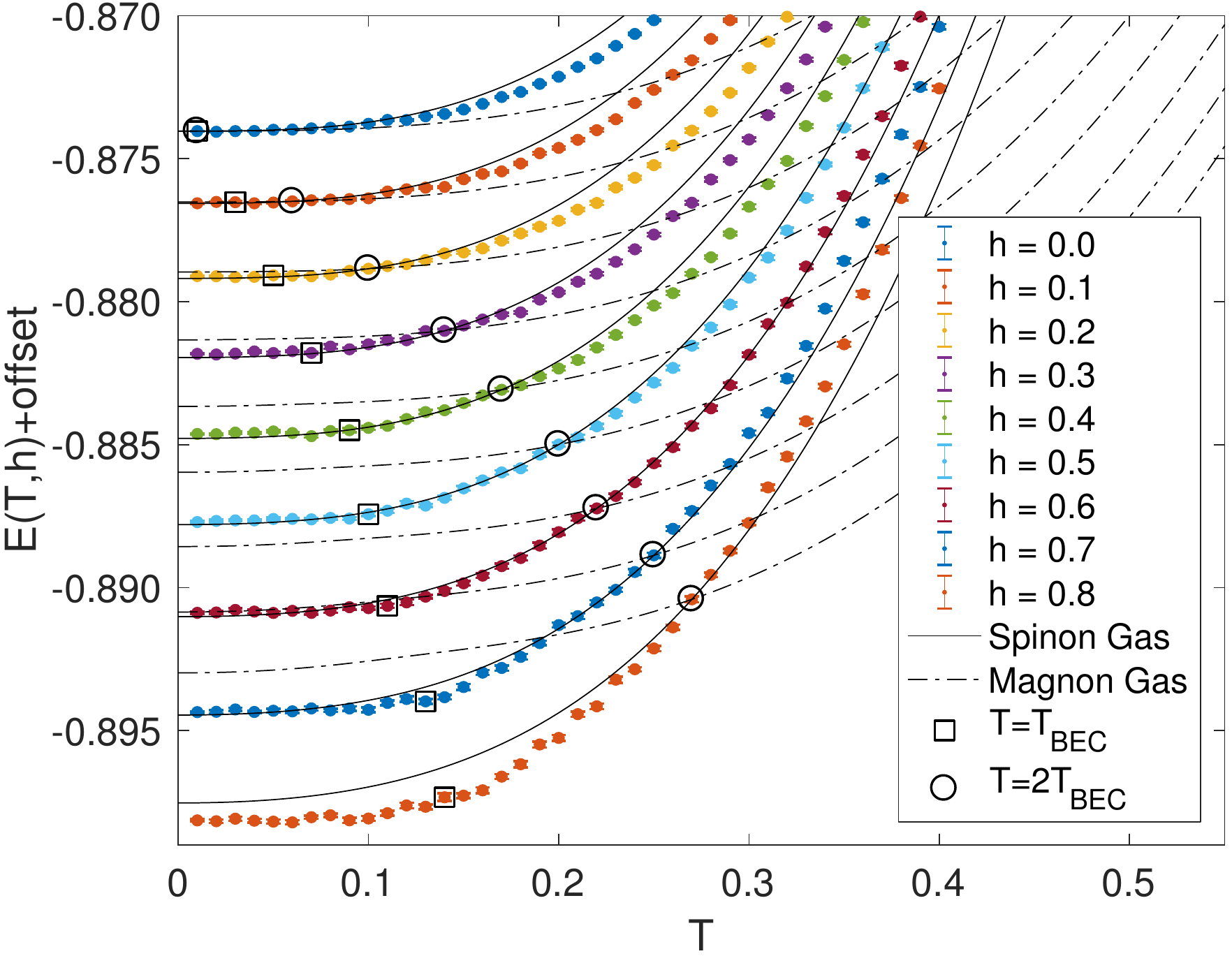}
\caption{Spinon gas for $T>T_{\rm BEC}(h)$. Colored $\bullet$ are QMC results for $E(T,h)+$offset (same data as \cref{f:bec}) compared to the field theory predictions for a gas of deconfined spinons (solid line) and a gas of magnons (broken line). The points $E(T_{\rm BEC},h)$, $E(2T_{\rm BEC},h)$ are marked with $\square$, $\circ$ respectively. Theory lines are numerically exact; QMC results' error bars are smaller than the markers. \label{f:sgas}}
\end{figure}

We calculate the thermal mass using a one-loop perturbation 
\begin{equation}
\Delta_T^2=\Delta_0^2(j)+ \Sigma(\Delta,h,T, k), \label{gapdef}
\end{equation}
where $\Sigma(\Delta,h,T, k)$ represents all one-loop corrections, to order $\alpha$ and $e^2$ (see \cref{f:loops}). 
At the DQC point ($j_c$) the $T=0$ spinon mass $\Delta_0(j)$ vanishes, so we set \mbox{$\Delta_0 = 0$} and numerically solve the transcendental equation for $\Delta_T$ [see \cref{gapansatz}]. 
The fine black lines in \cref{f:pd2} are field theory estimates of the crossover temperature, each obtained by solving the implicit equation for the spinon mass gap \mbox{$\Delta_T(h_c,T_c)-\mu h_c = \delta$} [\cref{gapdef}] for a different small value of $\delta\in\{10^{-6},10^{-4}\}$. 
Solving for $\delta=0$ is not possible due to the Mermin-Wagner theorem, but it is interesting to see that the curves depend only weakly on $\delta$ and fall close to the QMC results for the BKT transition---this supports the notion that the spinon gas should provide a good description of the lattice model above $T_{\rm BKT}$.

With these parameters established, we evaluate $E(h,T)$ for the spinon gas, accounting for all modes [\cref{eq:spinonGasModes}]\footnote{Note that for the spinon gas, the dispersions themselves depend on temperature through $\Delta_T$.}. 
In the relevant regime, gapless modes dominate $E(h,T)$.
Due to the gauge field, the spinons have an extra gapless mode $\omega_\gamma$, which is not present in the magnon description. 
Moreover, across the range of fields and temperatures \mbox{($T>T_{\rm BEC}$)} considered, the system remains close the transition \mbox{$(\Delta_T(h)-\mu h)\ll \mu h$}. 
Therefore there are {\it two} nearly-gapless spinon modes, $\omega_-\approx ck^2/(2\Delta_T) + \Delta_T-\mu h$; in contrast, a magnon gas has just {\it one} equivalent nearly-gapless mode [\cref{Cont_Zeeman}].
As a result, spinon and magnon theories will exhibit markedly different behavior for the statistical energy. 
This is indeed what we find. 

In \cref{f:sgas} we plot QMC results for $E(T,h)+F(h)$. 
Note that the QMC data and offset $F(h)$ in \cref{f:sgas} are identical to \cref{f:bec}, but here we are interested in testing the theories of magnon and spinon \textit{gases} at \textit{intermediate temperature}. 
As before, the theory has no remaining free parameters, so no fitting has been performed, but since the energy offset is not described by the field theory, we have shifted the theory curves so that they cross the QMC energy lines at $T=2T_{\rm BEC}(h)$. 
This choice of offset is somewhat arbitrary; we chose a simple assumption to make the analysis more clear, but our results do not depend on the exact choice of $T$.  
The spinon theory exhibits an excellent match to the numerical results, while the magnon theory is clearly incompatible. 
We therefore conclude that the gas phase of the system (above the BEC) \textit{cannot} be described in terms of conventional (nonfractional) magnetic excitations and the excitations are indeed spinons. 
This serves as additional evidence for our title claim: the BEC is formed from these same excitations, so the BEC must therefore be a BEC of spinons. 

In both \cref{f:bec,f:sgas} we find that the spinon theory works best for \textit{intermediate} fields. 
For small $h$ the density of spinons is low; the spinon contribution to the energy is therefore small and masked by other high-$T$ contributions not described by the theory. 
For large $h$, the system is too far from the DQC transition; this low-energy description becomes invalid and additional higher-order terms come into play. 
The highest field presented in \cref{f:bec,f:sgas}, $h=0.8$ is already nearly 50\% of the saturation field (and extremely high field for most materials) and the magnetization density would be a few percent \cite{iaizzi2018metamag}. 
At these densities, the spinon-spinon interactions may be more complex, and the low-energy theory we describe here may indeed no longer be accurate. 

%=================================================
\section{Conclusions \label{s:discussion}}

%\textit{Discussion.}---
We have studied deconfined quantum criticality in the presence of a magnetic field. 
The field dramatically alters the DQC point; 
breaking the global spin rotational symmetry, it unlocks a rich multicritical point, which is a complex intersection of N\'{e}el, VBS, and field-induced BEC phases. 
Our results provide evidence that the excitations in the BEC phase are indeed deconfined $S=1/2$ spinons, and not conventional $S=1$ magnons, thus extending the known DQC phenomenology. 
Our results join a mounting body of evidence \cite{sandvik2007,sandvik2011,suwa2016,shao2016,ma2018} that the transition between the N\'{e}el and VBS phases is indeed described by deconfined quantum criticality. 
In our case, we do this by directly interrogating the thermodynamic behavior of the excitations, rather than attempting to characterize the nature of the transition (continuous or weakly first-order). 
Moreover, we show that the emergent U(1) gauge field plays a critical role, contrary to expectations of Refs. \onlinecite{sandvik2011,scammell2015}.

This work considered only a small portion of this phase diagram near the DQC point separating the N\'{e}el and VBS states. 
Whether or not spinons remain deconfined along the extended quantum critical line of the N\'{e}el, VBS and BEC intersection is still an open question and warrants future non-perturbative studies.  
Combining our results with the previously-studied zero-field \cite{sandvik2011} and high-field \cite{iaizzi2018metamag,mythesis} cases, we were still unable to include even a schematic of the full $T$-$j$-$h$ phase diagram, in part because large system sizes (requiring long simulations) are needed to correctly extract the phase boundaries. 
This topic merits further exploration both numerically and theoretically. 

Beyond the DQC context in which we have developed our theory and simulations here, 
our work is also relevant to gapless spin liquid phases, which are the subject of active investigation both experimentally and theoretically \cite{song2018}. 
High-precision low-$T$ heat capacity studies of candidate gapless spin liquid materials would be the most natural way to test the BEC and spinon gas results we have presented here.

%=================================================
%\section*{Acknowledgements}
\begin{acknowledgments}
%\textit{Acknowledgements.}---
The work of A.I. and A.W.S. was supported by the NSF under Grant No.~DMR-1710170 and by a Simons Investigator Award; A.I. was also supported in part by the Ministry of Science and Technology (MOST) of Taiwan under Grant No.~107-2811-M-002-061. 
H.S. acknowledges support from the Australian-American Fulbright Commission.
The work of O.P.S. has been supported by Australian Research Council, Grant No.~DP160103630. 
H.S. and  O.P.S. also acknowledge support by the Australian Research Council Centre of Excellence in Future Low-Energy Electronics Technologies (CE170100039). 
The computational work reported in this paper was performed in part on the Shared Computing Cluster administered by Boston University's Research Computing Services. 
\end{acknowledgments}

%=================================================================================================
%=================================================
%=================================================
\appendix
%==================================================

%=================================================
\section{Fitting details} \label{supp:fittingDetails}
%=================================================
%\subsection{Fitting the phenomenological parameters}

In order to use field theory to make predictions, we must first determine the values of the various coupling ratios. 
Fitting the phenomenological field theory parameters, $\{c,\Delta_0, \alpha, e\}$, constitutes an important part of the present analysis. 
We will now describe how each parameter is obtained:\\
{\bf i)} The spinon speed is known from previous studies \cite{suwa2016} to be \mbox{$\frac{c}{J+Q} = 2.31(5)$}.
In our units ($J=0.045$, $Q=1$), $c = 2.42$; the spinon velocity is not expected to change due to finite field or finite temperature effects.\\ 
{\bf ii)} The spinon mass $\Delta_0^2\propto j-j_c$, but the QMC is taken at the QCP ($j=j_c$) and hence $\Delta_0=0$. \\
{\bf iii)} Since $\Delta_0=0$, the condensate energy [\cref{eq:condensate}] is given purely in terms of $\mu h$ and $\alpha=\mu h \tilde{\alpha}$. 
Comparing with QMC data, we find $\tilde{\alpha}=\frac{2}{3}\pi c^2 (0.32)$ [see \cref{Fparams}].\\
{\bf iv)} Using magnetic susceptibility QMC data \cite{sandvik2011} and this QFT prediction, 
\begin{align}
\label{ChiT}
\chi/T=\mu^2\frac{1}{2\pi c^2}\left(\frac{\Delta}{T}\frac{1}{1-e^{-\frac{\Delta}{T}}}-\ln\left(e^{\frac{\Delta}{T}}-1\right)\right),
\end{align}
we find [\cref{Fparams}(a)] that the linear approximation \mbox{$\Delta(h=0,T)=\Theta T$}, where \mbox{$\Theta\approx0.59$} quantitatively fits the data. 
From this we determine $\tilde{e}=0.75 c$ (from here on we take $c$ as dimensionless).

%=================================================
\section{Renormalization details}\label{supp:renormDetails}
%=================================================
\subsection{Mass renormalization}

The zero-temperature spinon mass vanishes \mbox{$\Delta_0(j_c)=0$} because the system is tuned to the critical point \mbox{$m_0^2=\mu^2 h^2$}. 
However, at nonzero temperatures the spinons acquire mass due to interactions with the heat bath. 
We obtain this mass correction from the one-loop correction to the spinon propagator:
\begin{align}
\label{Gspinon}
D_\sigma(p_0,\bm p)=& \frac{i}{(p_0+\sigma \mu h)^2-{\bm p}^2 - m_0^2+i\epsilon} \notag \\
&  \to \frac{i}{(p_0+\sigma \mu h)^2-{\bm p}^2 - m_0^2 - \Sigma}, \\
\Delta^2(p,h,T)\equiv& m_0^2+\Re\Sigma = \Re\Sigma_1 + \Sigma_2 + \Sigma_3, \label{Gspinon2}
\end{align}
where $\Sigma_i$ refer to loop corrections with notation defined in \cref{f:loops}(a) and $\Re$ denotes real part. 

Implicit in \cref{Gspinon,Gspinon2} is integration over internal gauge field propagators; we must therefore evaluate loop corrections to these propagators.
We work in the Coulomb gauge, and upon inclusion of the one-loop corrections the propagators become
\begin{align}
\label{Gcoulomb}
G^{00}(p_0,\bm p)&=\frac{i}{{\bm p}^2 + i\epsilon}\to\frac{i}{\bm p^2+\Pi^{00}} ,\\
\label{Gmagnetic}
G^{ij}(p_0,\bm p)&=i\frac{\delta^{ij}-p^ip^j/{\bm p}^2}{p_0^2-{\bm p}^2 + i\epsilon}\to i\frac{\delta^{ij}-p^ip^j/{\bm p}^2}{p_0^2-{\bm p}^2 - \Pi^{ij}},
\end{align}
where $\Pi^{\mu\nu}=\Pi^{\mu\nu}_1 + \Pi^{\mu\nu}_2$, with $\Pi^{\mu\nu}_1, \Pi^{\mu\nu}_2$ as depicted in \cref{f:loops}(b). 

We explicitly consider the renormalization due to nonzero $T$; all purely quantum corrections (i.e. ultraviolet divergences) are implicitly taken care of by absorbing them into redefinitions of the Lagrangian coupling constants at $T=0$. 
We do not consider ultraviolet renormalization any further; interested readers should consult any standard textbook on scalar QED, e.g. Ref. \onlinecite{Srednicki}.

%=================================================
\begin{widetext}
%=================================================

%%%%%%%%%%%%%%%%%
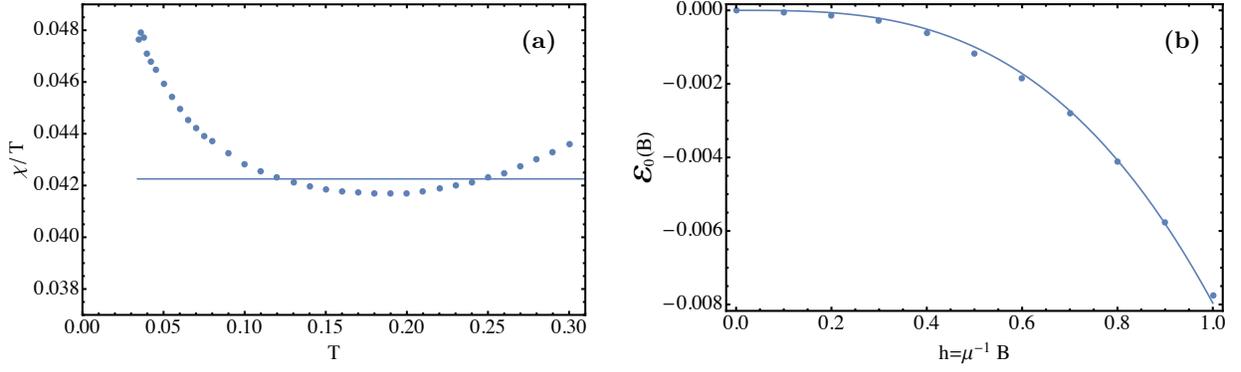
\begin{figure}
 {\includegraphics[width=0.43\textwidth,clip]{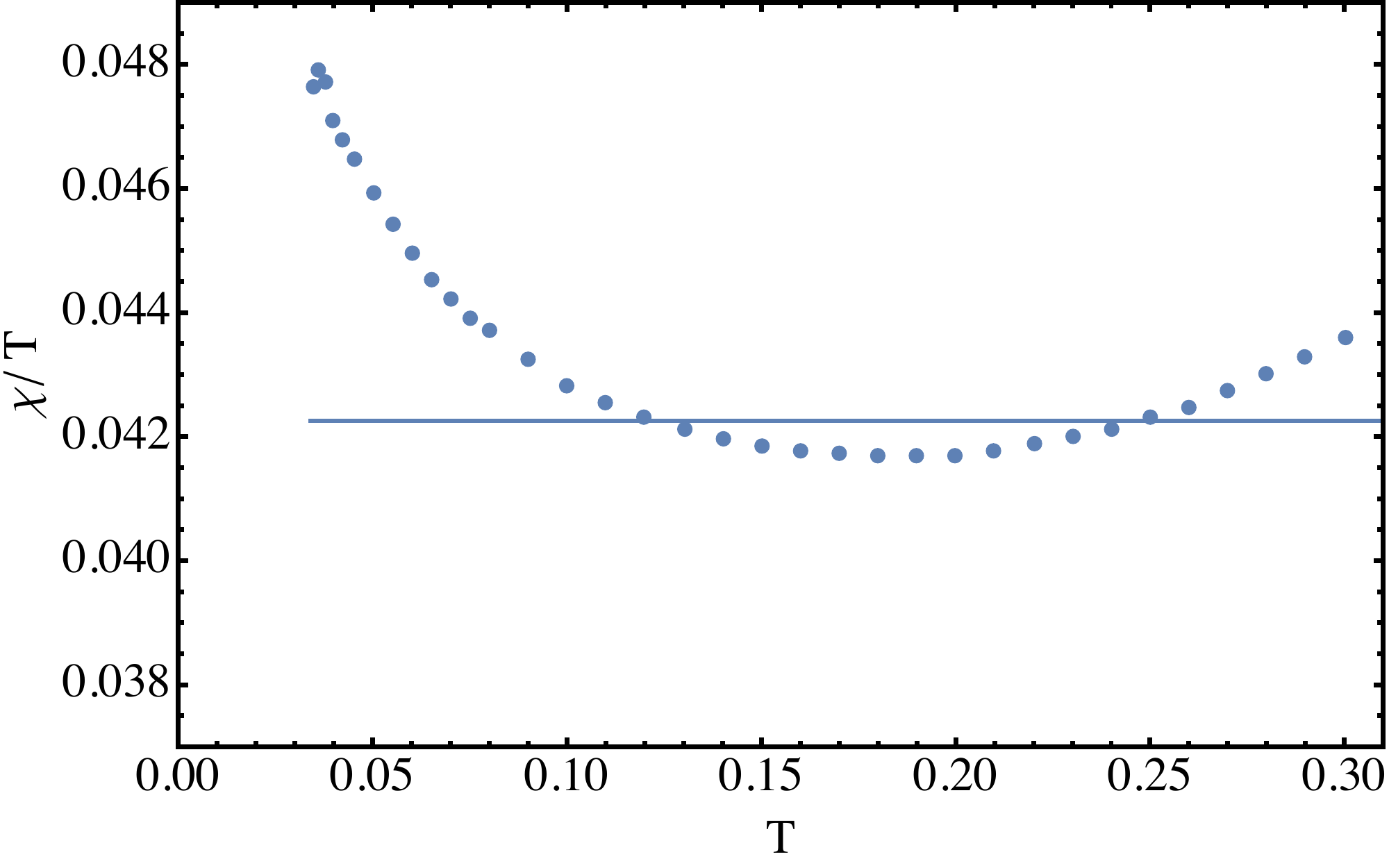}}\hspace{0.5cm}
 {\includegraphics[width=0.44\textwidth,clip]{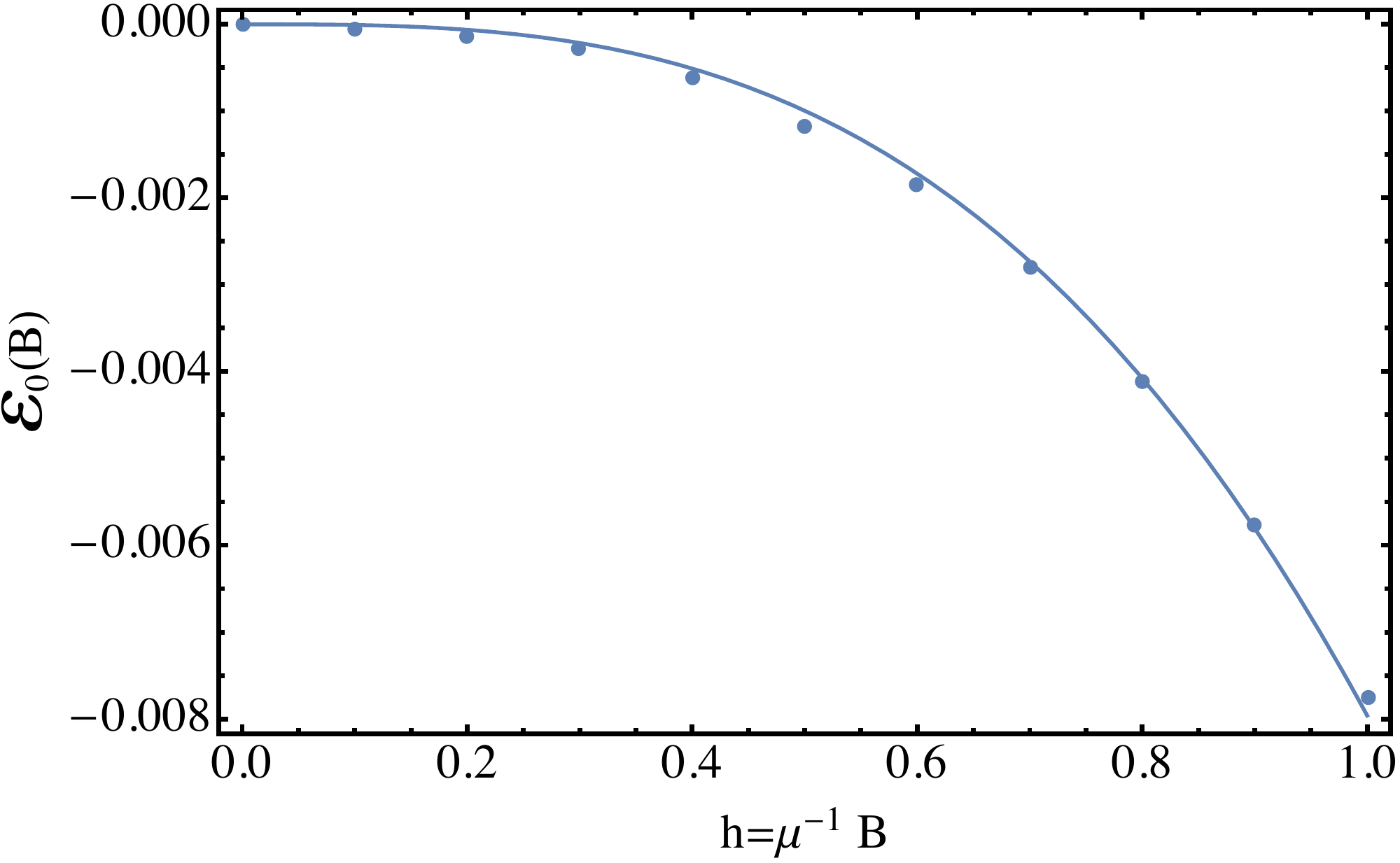}}
 %add labels (a), (b) to panels
 \begin{picture}(0,0) 
	\put(-270,120){\textbf{(a)}} %coordinates in points from lower right corner of fig 
	\put(-27,120){\textbf{(b)}}
 \end{picture}
 \caption{Fitting phenomenological parameters. (a) Magnetic susceptibility over temperature, $\chi/T$: Points are QMC data from Ref.~\onlinecite{sandvik2011}; the line is \cref{ChiT} taking $\Delta(h=0,T)=\Theta T$, with $\Theta=0.59$ (which produces a constant). (b) Condensate energy vs. field at $T=0$, ${\cal E}_0(h)$: Points are our QMC results; the line is a QFT prediction from \cref{eq:condensate} using parameters $\{ \tilde\alpha, c\}=\{\frac{2}{3}\pi c^2 (0.32), 2.42\}$. Everywhere we take $Q=1$.}
\label{Fparams}
\end{figure}
%%%%%%%%%%%%%%%%%

%=================================================
\subsection{Loop integrals at $T,h\neq0$\label{loopintegrals}}

%%%%%%%%%%%%
\begin{figure} 
\includegraphics[width=150mm]{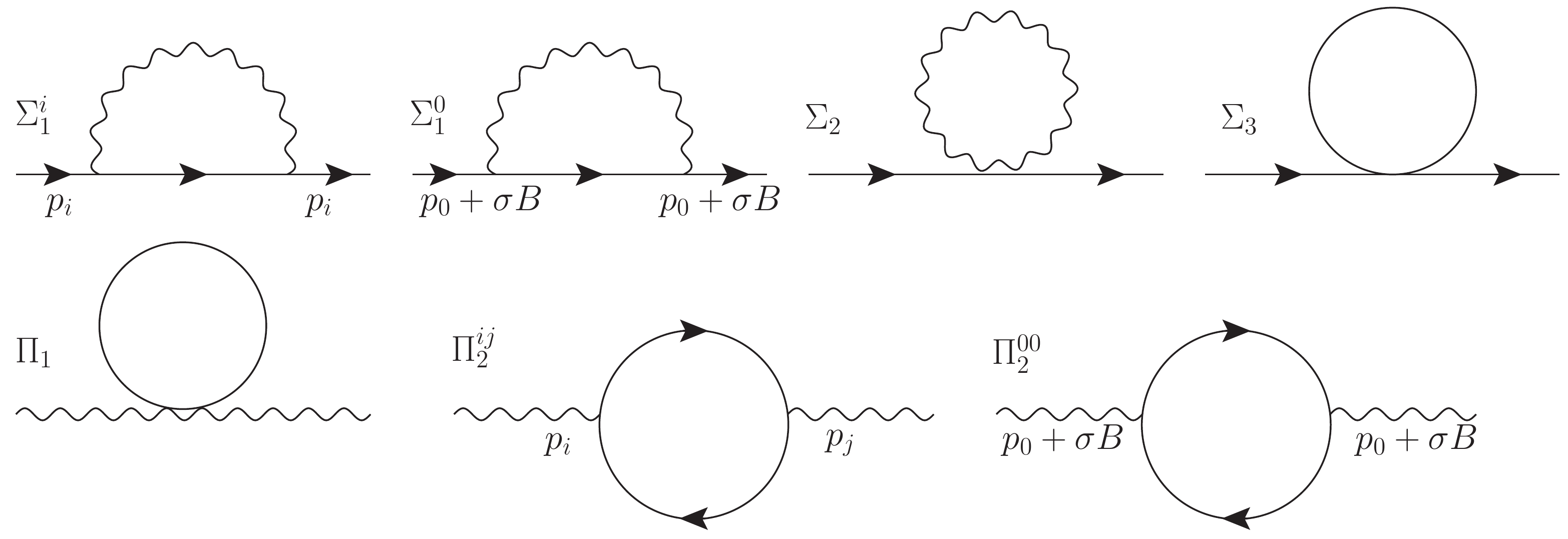}
\caption{Loop corrections to (a) the spinon Green's function \cref{Gspinon}, and (b) the gauge field Green's functions \cref{Gcoulomb,Gmagnetic}. Labeling of external momenta is used to help define the various self energy components. Terms $\Pi_2^{0i}$ are intentionally omitted. In both figures we have used $B\equiv \mu h$.
\label{f:loops}}
\begin{picture}(0,0) 
\put(-240,150){\text{{(a)}}} 
\put(-240,85){\text{{(b)}}} 
\end{picture}
\end{figure}
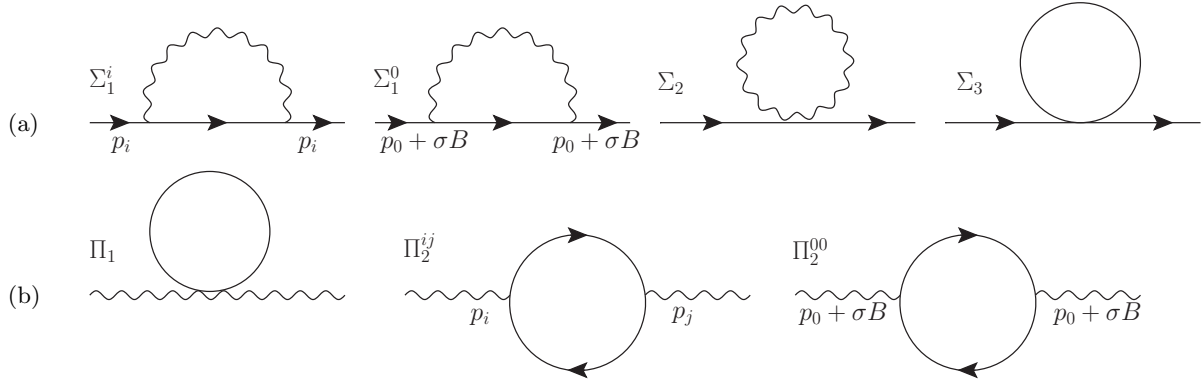
%%%%%%%%%

We now evaluate the nonzero $T,h$ contributions to the loop integrals.  
Consider the first diagram of \mbox{\cref{f:loops}(a) $(\Sigma_1)$.}  
Evaluating this diagram with an internal Coulomb line (i.e. using $G^{00}$) gives
\begin{align}
\Sigma^{0}_1(p,h,T)&=-S_1 e^2 \int \frac{d^2l}{(2\pi c)^2}\frac{1}{(|\bm l-\bm p|^2 +\Pi^{00}) 2\omega_0}\left\{\omega_{0}^2 n(\omega_{+}) + \omega_{0}^2n_{-}(\omega_{-})\right\} \ \ \ \text{$+$ quantum correction},
\end{align}
with $\omega_\pm=\omega_0\pm  \mu h$ and $\omega_0=\sqrt{l^2+\Delta^2}$; $n(\omega)$ is the usual Bose factor. 
We have also explicitly taken the on-shell condition $p_0=0$ for the Coulomb field. 
The prefactor $S_1=1$ is a combinatorial factor of the loop diagram.  
Note: at $T=0$ the magnetic field can be `gauged out', and so we expect the loops to be independent of $h$ in this limit. 
Evaluating this diagram with an internal transverse field line (i.e. internal lines correspond to $G^{ii}$) gives
\begin{align}
\label{Principle}
\Sigma^{i(-)}_1=-S_1 e^2 \int \frac{d^2l}{(2\pi c)^2}\frac{4p^2\sin^2\theta}{2\omega_0 2\omega_2}\Big\{&
\left[(1+n_{-})(1+n_2)-n_- n_2\right]\frac{1}{p_0-\omega_--\omega_2+i\epsilon} \\
\notag - &\left[n_- (1+n_2)-(1+n_{-})n_2\right]\frac{1}{p_0-\omega_-+\omega_2+i\epsilon}\\
\notag-&\left[(1+n_+) n_2-n_{+}(1+n_2)\right]\frac{1}{p_0+\omega_+-\omega_2+i\epsilon}\\
\notag-&\left[(1+n_{+})(1+n_2)-n_+ n_2\right]\frac{1}{p_0+\omega_++\omega_2+i\epsilon}\Big\}.
\end{align} 
Again we use $\omega_0=\sqrt{l^2+\Delta^2}, \omega_\pm=\omega_0\pm \mu h$, and also define $\omega_2=|\bm l-\bm p|$. 
The notation $\Sigma^{i(\pm)}_1$ refers to the dispersion of the external spinon, $\omega(p)_\pm=\sqrt{p^2+\Delta^2}\pm \mu h$, and $\Sigma^{i(+)}_1(h)=\Sigma^{i(-)}_1(-h)$.  
When the external momentum is taken on-shell, i.e. $p_0=\sqrt{\bm p^2+\Delta^2}+\sigma \mu h$, then at $T=0$ we find that $\Sigma^{i(+)}_1(h)=\Sigma^{i(-)}_1(h)$, which is reminiscent of `gauging out' $h$ at $T=0$.

Now consider the second diagram of \cref{f:loops}(a) $(\Sigma_2)$. 
Evaluating this diagram with an internal Coulomb line ($G^{00}$), and a transverse gauge field line ($G^{ii}$) gives 
\begin{subequations}
\begin{align}
\Sigma^{0}_2&=0, \\
\Sigma^{i}_2&=S_2 e^2 \int_{\rm IR} \frac{d^2l}{(2\pi c)^2}\frac{n(l)}{l}, 
\end{align}
\end{subequations}
respectively. 
Here, the combinatorial factor is $S_2=2$. 
The subscript IR indicates that we used an infrared cutoff to tame the divergences of this integral. 
For this purpose, we take the natural infrared energy scale to be the spinon mass, $\Delta$. 
For a more sophisticated treatment of infrared divergences in non-zero temperature scalar QED we refer the reader to Ref.~\onlinecite{blaizot1996}. 
Finally, consider the third diagram of \cref{f:loops}(a) $(\Sigma_3)$. 
Here we integrate over the internal spinon propagator, which gives 
\begin{align}
\Sigma_3&=S_3 \alpha\int \frac{d^2l}{(2\pi c)^2}\frac{1}{2\omega_0}\left(n(\omega^+_l) + n(\omega^-_l)\right).
\end{align}
Again, $\omega^\pm_l=\omega_0(l)\pm \mu h$ and $\omega_0(l)=\sqrt{l^2+\Delta^2}$.
The combinatorial factor is $S_3=(N+2)/2=3$, where $N=4$.

Let us now consider the loop diagrams appearing in \cref{f:loops}(b). 
The first diagram is evaluated simply, and there is no need to separate the components of the gauge field (considering only the thermal contribution) 
\begin{align}
\Pi^{\mu\nu}_1&=C_1 g_{\mu\nu} e^2\int \frac{d^2l}{(2\pi c)^2}\frac{1}{2\omega_0}\left(n(\omega^+_l) + n(\omega^-_l)\right).
\end{align}
The combinatorial factor is $C_1=N=4$. 
In the second diagram, we once again separate different gauge field components. 
For an external Coulomb field, and using the on-shell condition $p_0=0$, we find
\begin{align}
\Pi^{00}_2(\bm p, h)=&\frac{C_2 e^2}{(2\pi c)^2}\int \frac{d^2l  \omega_0(l)}{{4\omega_0(p)}} \times\\
\notag&\left\{\left[\frac{1+n(\omega^-_l)+n(\omega^-_p)}{\omega^-_l  + \omega^-_p+i\epsilon}- \frac{n(\omega^-_l)-n(\omega^+_p)}{\omega^-_l  - \omega^+_p+i\epsilon}\right]  +\left[\frac{1+n(\omega^+_l)+n(\omega^+_p)}{\omega^+_l  + \omega^+_p +i\epsilon}- \frac{n(\omega^+_l)-n(\omega^-_p)}{\omega^+_l  - \omega^-_p +i\epsilon}\right] \right\},
\end{align}
with $\omega^\pm_p=\omega_0(p)\pm \mu h$ and $\omega_0(p)=\sqrt{p^2+\Delta^2}$. 
The combinatorial factor $C_2=N/2=2$. 
For an external transverse gauge field, and using $p_0=|\bm p|$, we find
\begin{align}
\Pi^{ii}_2(\bm p, h)=&\frac{C_2 e^2}{(2\pi c)^2} \int \frac{d^2l}{{2\omega_0(l) 2\omega_0(p)}} (\bm l+\bm p)^2 \times\\
\notag& \left\{\left[\frac{1+n(\omega^-_l)+n(\omega^-_p)}{|\bm p|-\omega^-_l  - \omega^-_p}+ \frac{n(\omega^-_l)-n(\omega^+_p)}{|\bm p|-\omega^-_l + \omega^+_p}\right] + \left[\frac{-(1+n(\omega^+_l)+n(\omega^+_p))}{|\bm p|+\omega^+_l  + \omega^+_p}+ \frac{-(n(\omega^+_l)-n(\omega^-_p))}{|\bm p|+\omega^+_l  - \omega^-_p}\right] \right\}.
\end{align}

%=================================================
\section{Total energy and the partition function} \label{supp:totEn}
%=================================================
\subsection{Effective Lagrangian}

In the spinon gas phase, the Lagrangian is 
\begin{align}
\label{Lza}
{\cal L}[z,a_\mu]&=(D_\mu z)^\dag(D^\mu z)+\mu h z^\dag\sigma_3 D_0 z - \mu h (D_0 z)^\dag\sigma_3z - (\Delta_0^2 - \mu^2 h^2) z^\dag z - \frac{1}{2}\alpha(z^\dag z)^2 - \frac{1}{4}f_{\mu\nu}f^{\mu\nu},
\end{align}
with $D_\mu=\partial_\mu - iea_\mu$. 
The spinon condensate occurs by tuning $\mu^2 h^2> \Delta_0^2$, for which a non-zero expectation $\braket{z}\equiv z_0$ develops. 
For our purposes, the zero-temperature spinon mass parameter is zero, i.e. $\Delta_0=0$, therefore the application of any non-zero magnetic field will result in spinon condensation. 
However, we wish to consider non-zero temperatures, such that a thermal spinon mass $\Delta(T)>\Delta_0$ is induced via interactions (i.e. the loop corrections in \cref{f:loops}). 
To consistently discuss the disordered phase, we must include the thermal spinon mass in the Lagrangian such that $\Delta^2(T)\geq \mu^2 h^2$, otherwise the Lagrangian will describe the wrong ground state (the spinon BEC) and the fluctuations thereof. 
This amounts to performing a reorganization of the perturbation theory, which we call the effective Lagrangian approach. 

To illustrate our effective Lagrangian approach, which is employed to overcome the expansion about the wrong ground state, we consider the Lagrangian in absence of the magnetic field (purely to facilitate the presentation). 
We also include the expansion of terms up to $e^2$ which can be easily reproduced from the partition function (but we do not present those details). 
We find
\begin{align}
\label{effective}
\notag{\cal L}&=\partial_\mu z^\dag\partial^\mu z +iea_\mu z^\dag\partial_\mu z - iea^\mu\partial_\mu z^\dag z - \frac{1}{4}f_{\mu\nu}f^{\mu\nu} - \Delta_0^2 z^\dag z - \frac{1}{2}\alpha(z^\dag z)^- e^2a_\mu a^\mu z^\dag z+ e^2 z^\dag a^\mu \partial_\mu z \ a^\nu\partial_\nu z^\dag z,\\
{\cal L}&\equiv{\cal L}_K - \Delta_0^2 z^\dag z - \frac{1}{2}\alpha(z^\dag z)^2  - e^2a_\mu a^\mu z^\dag z+ e^2 z^\dag a^\mu \partial_\mu z \ a^\nu\partial_\nu z^\dag z,
\end{align}
which we reorganize as
\begin{align}
\notag{\cal L}=&{\cal L}_K - \left[\Delta_0^2-S_1e^2\braket{a^\mu \partial_\mu z \ a^\nu\partial_\nu z^\dag}+S_2e^2\braket{a_\mu a^\mu}+S_3\alpha\braket{z^\dag z}\right] z^\dag z - \frac{1}{2}\alpha\left[(z^\dag z)^2- 2S_3\braket{z^\dag z}z^\dag z \right] \\*
\notag& \hspace{0.65cm} -  e^2\left[C_1\braket{z^\dag z}- C_2\braket{z^\dag\partial_\nu z \ z^\dag\partial^\nu z}\right]a_\mu a^\mu- e^2\left[a_\mu a^\mu z^\dag z - S_2\braket{a_\mu a^\mu}z^\dag z - C_1\braket{z^\dag z} a_\mu a^\mu \right]\\*
\notag &
 \hspace{0.65cm} + e^2 \left[z^\dag a^\mu \partial_\mu z \ a^\nu\partial_\nu z^\dag z  + S_1 \braket{a_\nu a^\mu \partial^\nu z \partial_\mu z^\dag} z^\dag  z + C_2 \braket{z^\dag\partial^\nu z \ z^\dag\partial_\mu z}a_\nu a^\mu \right],\\
\label{reorganized}
{\cal L}=&{\cal L}_K -\left[\Delta_0^2+\Sigma_1+\Sigma_2+\Sigma_3\right]z^\dag z - \left[\Pi^{\mu\nu}_1+\Pi^{\mu\nu}_2\right]a_\nu a^\mu - \left[\frac{1}{2}\alpha(z^\dag z)^2- \Sigma_3 z^\dag z \right] \\*
\notag& \hspace{0.65cm} - \left[e^2a_\mu a^\mu z^\dag z - \Sigma_2 z^\dag z - \Pi_1^{\mu\nu} a_\nu a^\mu \right] + \left[e^2 z^\dag a^\mu \partial_\mu z a^\nu\partial_\nu z^\dag z  + \Sigma_1 z^\dag z + \Pi_2^{\mu\nu} a_\nu a^\mu \right].
\end{align}
Here the bracket notation, e.g. $\braket{z^\dag z}$, implies loop integration over the fields inside. 
The Lagrangian ${\cal L}_K$ stands for the purely {\it kinetic} parts. 
This reorganization is {\it exact}. 
Finally, we reach the key point: the effective Lagrangian is given by 
\begin{align}
\label{bilinear}
{\cal L}_E&={\cal L}_K -\left[\Delta_0^2+\Sigma_1+\Sigma_2+\Sigma_3\right]z^\dag z - \left[\Pi^{\mu\nu}_1+\Pi^{\mu\nu}_2\right]a_\nu a^\mu. 
\end{align}
${\cal L}_E$ is purely bilinear in all dynamic field variables. 
Note: loops such as $\braket{z^\dag z}$ are no longer dynamical variables, just numbers. 
In the main text, the quasiparticle dispersions are obtained directly from ${\cal L}_E$ in the disordered spin gas phase, and importantly this method is entirely equivalent to the normal Dyson summation of loop corrections to the Green's functions. 

%=================================================
\subsection{Total energy}

The bilinear effective Lagrangian ${\cal L}_E$ [\cref{bilinear}] is a consistent means to obtain the quasiparticle dispersions. 
However, when one wishes to consider the {\it total energy} of the system, simply summing over all modes with renormalized dispersion obtained from ${\cal L}_E$ [\cref{bilinear}] is {\it not} equivalent to the standard perturbative expansion of the partition function. 
One must still perform a perturbative expansion, but now in the shifted interaction Lagrangian: 
\begin{align}
\label{reorganizedInteraction}
&{\cal L}_I=- \left[\frac{1}{2}\alpha(z^\dag z)^2- \Sigma_3 z^\dag z \right] - \left[e^2a_\mu a^\mu z^\dag z - \Sigma_2 z^\dag z - \Pi_1^{\mu\nu} a_\nu a^\mu \right] + \left[e^2 z^\dag a^\mu \partial_\mu z a^\nu\partial_\nu z^\dag z  + \Sigma_1 z^\dag z + \Pi_2^{\mu\nu} a_\nu a^\mu \right].
\end{align}
These terms contribute to the partition function expansion to the same order in the coupling constants, i.e. $\alpha$ and $e^2$.
%%%%%%%%%%%%%%%
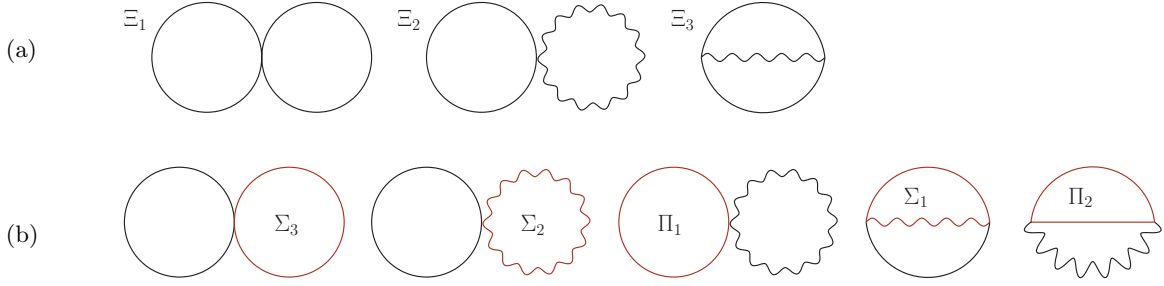
\begin{figure} 
\includegraphics[width=140mm]{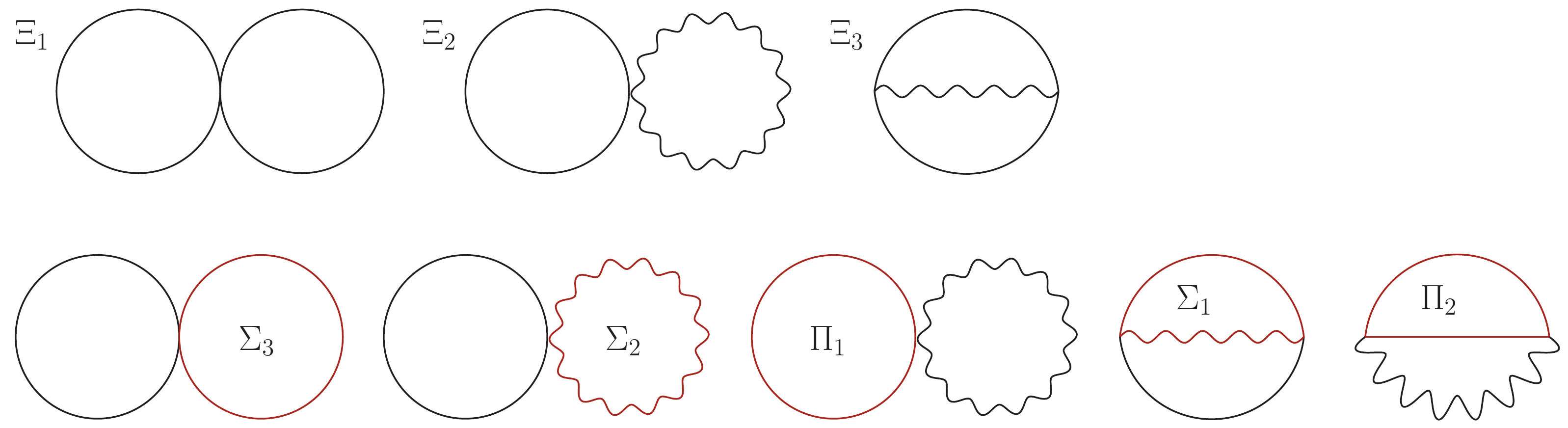}
\caption{Loop corrections to the partition function $\ln Z$ [\cref{partition}] due to the interaction terms in $\ln Z_I$. (a) The standard partition function loop corrections with interactions taken from \cref{effective}. (b) The additional corrections due to the reorganization of the perturbation theory, \cref{reorganized}. Here the loops traced by red lines are the self-energy loops appearing in the reorganized interaction Lagrangian, \cref{reorganizedInteraction}. 
\label{f:partitionloops}}
\begin{picture}(0,0) 
\put(-240,150){\text{{(a)}}} 
\put(-240,80){\text{{(b)}}} 
\end{picture}
\end{figure}
%%%%%%%
All in all, the partition function is expanded in the usual way,
\begin{align}
\label{partition}
\ln Z = \ln Z_E + \ln Z_I.
\end{align}
$Z_E$ is the partition function with ${\cal L}_E$ [\cref{bilinear}], which is straightforward to evaluate since it is bilinear in the fields, 
\begin{align}
\ln Z_E &= -V\sum_{i=1}^5\int \frac{d^2k}{(2\pi)^2}\ln\left(1-e^\frac{{-\omega^i_{k}}}{T}\right),
\end{align}
where the index $i$ labels the five different modes. 
On the other hand, $Z_I$ contains the interactions ${\cal L}_I$ [\cref{reorganizedInteraction}], and cannot be evaluated exactly. 
Instead, we use the usual perturbative expansion 
\begin{align}
\ln Z_I &= \frac{1}{Z_E}\int[Dz^\dag][Dz][Da_\mu] e^{\int{\cal L}_E} {\cal L}_I\equiv\braket{{\cal L}_I}.
\end{align}
Upon substituting the reorganized interaction \cref{reorganizedInteraction}, we find
\begin{align}
\label{lnZ}
\notag \frac{\ln Z_I}{\beta V} &=\frac{1}{2}\alpha\braket{(z^\dag z)^2} + e^2\braket{a_\mu a^\mu z^\dag z} - e^2 \braket{z^\dag a^\mu \partial_\mu z a^\nu\partial_\nu z^\dag z} -
\Sigma_3 \braket{z^\dag z} -  \Sigma_2 \braket{z^\dag z} - \Pi_1^{\mu\nu} \braket{a_\nu a^\mu}    + \Sigma_1 \braket{z^\dag z} + \Pi_2^{\mu\nu} \braket{a_\nu a^\mu},\\
&=-\frac{1}{2}\alpha\braket{(z^\dag z)^2} - e^2\braket{a_\mu a^\mu z^\dag z} + e^2 \braket{z^\dag a^\mu \partial_\mu z a^\nu\partial_\nu z^\dag z} \equiv -(\Xi_1 + \Xi_2 + \Xi_3),
\end{align}
where $\beta$ and $V$ are the inverse temperature and spatial volume. 
The top line of \cref{lnZ} contains all diagrams shown in \cref{f:partitionloops}: the first three terms on the RHS correspond to the diagrams of \cref{f:partitionloops}(a); the next five terms on the RHS correspond to the diagrams of \cref{f:partitionloops}(b). 
However, we see from the bottom line of \cref{lnZ} that after cancelation we are left with just the negative sum of the diagrams of \cref{f:partitionloops}(a). 

An expression for all loop diagrams in \cref{f:partitionloops} can be deduced from our expressions in \cref{loopintegrals}:
\begin{align}
\frac{\Xi_1}{\beta V}&=  \frac{1}{2}\alpha(N+2)\left[\int \frac{d^2l}{(2\pi c)^2}\frac{1}{2\omega_0}(n(\omega^+_l) + n(\omega^-_l))\right]^2,\\
\frac{\Xi_2}{\beta V}&=  \frac{1}{2}e^2N\left[\int \frac{d^2l}{(2\pi c)^2}\frac{1}{2\omega_0}(n(\omega^+_l) + n(\omega^-_l))\right]\left[\int_{IR} \frac{d^2l}{(2\pi c)^2}\frac{n(l)}{l}\right],
\end{align}
which are straightforward multiplicative loops. 
However, for $\Xi_3$ we must perform a non-trivial contour integration: 
\begin{align}
\label{Xi3}
\frac{\Xi_3}{\beta V}=S_1 e^2 &\int \frac{d^2l d^2p}{(2\pi c)^2}\frac{4p^2\sin^2\theta}{2\omega_p 2\omega_l 2\omega_{l-p}}\int^\beta_0d\tau\Big\{ \left[(1+n(\omega_p^-))e^{-\omega_p^-\tau}+n(\omega_p^+)e^{\omega_p^+\tau}\right] \times\\*
\notag&\left[(1+n(\omega_l^-))e^{-\omega_l^-\tau}+n(\omega_l^+)e^{\omega_l^+\tau}\right]\left[(1+n(\omega^\gamma_{l-p}))e^{-\omega^\gamma_{l-p}\tau}+n(\omega^\gamma_{l-p})e^{\omega^\gamma_{l-p}\tau}\right] \Big\}.
\end{align} 
Here $\omega^\pm_p=\omega_0(p)\pm \mu h$ and $\omega_0(p)=\sqrt{p^2+\Delta^2}$, while $\omega^\gamma_p=p$ (with $c=1$) is the dispersion of the U(1) gauge field. 
Integration over $\tau$ is trivial and can be performed analytically. 
However, this leads to a more complicated final expression, so we opt to keep the integral in this form.

Finally, the total energy is given by 
\begin{align}
\label{totalenergy}
E&=-T^2\partial_T\ln Z= V \sum_{i=1}^5\int \frac{d^2k}{(2\pi)^2}\left\{\omega^i_{k}n(\omega^i_{k})-Tn(\omega^i_{k})\partial_T\omega^i_{k}\right\} - T^2\partial_T \ln Z_I.
\end{align}
We use the formulation \cref{totalenergy} to obtain the results in the main text.

%=================================================
\subsection{Reproducing $E(T)$ predictions}

We provide a simple means for the interested reader to reproduce our theory predictions from \cref{f:bec,f:sgas}. 
Both are obtained by inserting the spinon and gauge dispersions into \cref{totalenergy}, which will be a different procedure for each phase. 
For the BEC phase, we calculate the dispersions explicitly at $T=0$ which means that the dispersions are independent of $T$ and so $\partial_T$-terms appearing in \cref{totalenergy} vanish. 
Substituting the dispersions from \cref{BECspec}, \cref{f:bec} immediately follows. 

For the spinon gas phase, we explicitly work at $T>0$, which has two main effects: (i) it dramatically renormalizes the spinon mass as well as the screening of the {\it Coulomb} component of the gauge field $a_0$. 
(ii) All $\partial_T$-terms appearing in \cref{totalenergy} must be evaluated, as they give important corrections. 
Part (i) is a complicated procedure which amounts to self-consistently solving for the spinon gap taking into account the loops in \cref{f:loops}. 
The numerical solution for \mbox{$\Delta(j_c,h,T,k=0)$} is well approximated by the following phenomenological ansatz: 
\begin{align}
\label{gapansatz}
\Delta^2(h,T)&=\sqrt{\Theta^2 T^2 e^{- \gamma \frac{\mu^2 h^2}{T^2}}+\mu^2 h^2 },
\end{align}
where $\gamma$ and $\Theta$ are dimensionless constants determined by fitting to be $\gamma=1.32$ and $\Theta=0.59$. 
Hence, substituting the gap ansatz \cref{gapansatz}, along with the dispersions in the spinon gas phase \cref{eq:spinonGasModes}, into \cref{totalenergy} [making use of Eqs. (\ref{lnZ})--(\ref{Xi3})] allows one to reproduce the curves in \cref{f:sgas}.

%=================================================
\section{Symmetry-breaking mechanism and gauge field mass generation}\label{supp:symmetry}
%=================================================
\subsection{Global symmetry-breaking patterns and Goldstone counting}

To discuss the global symmetries, the explicit and spontaneous symmetry breaking, and the expected number of Goldstone modes, it is illuminating to first consider the field theory without any coupling to the U(1) gauge field (we will reintroduce the gauge field later). 
The Lagrangian reads
\begin{align}
\label{Lz}
{\cal L}[z]&=(\partial_\mu z)^\dag(\partial^\mu z)+\mu h z^\dag\sigma_3 \partial_0 z - \mu h (\partial_0 z)^\dag\sigma_3z - (\Delta_0^2-\mu^2 h^2) z^\dag z - \frac{1}{2}\alpha(z^\dag z)^2,
\end{align}
where $z=(z_1,z_2)^T$, and we take $\Delta_0^2>0$ throughout this section.

To understand the global symmetries in the broken and unbroken phases, we introduce a matrix field 
\begin{align}
\Phi=\frac{1}{\sqrt{2}}\begin{pmatrix} z_2^* & z_1 \\ -z_1^* & z_2 \end{pmatrix},
\end{align}
and rewrite \cref{Lz} in terms of $\Phi$:
\begin{align}
\label{L_Phi}
\notag{\cal L}[\Phi]&=\Tr\left[(\partial_\mu \Phi)^\dag(\partial^\mu \Phi)\right]+\mu h \Tr\left[\Phi^\dag\sigma_3 (\partial_0 \Phi)\right] - \mu h \Tr\left[(\partial_0 \Phi)^\dag\sigma_3\Phi\right] - (\Delta_0^2-\mu^2 h^2)\Tr\left[\Phi^\dag\Phi\right]-\frac{1}{2}\alpha\Tr\left[\Phi^\dag\Phi\right]^2.
\end{align}
Let us now analyze the global symmetries: \\

\noindent $\bullet$ If the external field is set to zero $(h=0)$ the Lagrangian is invariant under two separate SU(2) transformations: the left and right multiplication by SU(2) matrices $U_L$ and $U_R$,  such that  ${\cal L}[U_L\Phi]={\cal L}[\Phi]$ and ${\cal L}[\Phi U_R]={\cal L}[\Phi]$. 
The global symmetry is therefore $\rm SU(2)_L\times SU(2)_R$. \\

\noindent $\bullet$ Turning on the external field, $h\neq0$, we need to check the transformation properties of the terms linear in $h$ in \cref{L_Phi}. 
An explicit calculation shows that under separate left and right multiplication these terms transform as
\begin{align}
\Tr\left[(U_L\Phi)^\dag\sigma_3 (\partial_0 U_L\Phi)\right]&=\Tr\left[\Phi^\dag(\sigma_3 + U_L^\dag[\sigma_3,U_L])(\partial_0 \Phi)\right],\\
\Tr\left[(\Phi U_R)^\dag\sigma_3 (\partial_0\Phi U_R)\right]&=\Tr\left[U_R^\dag\Phi^\dag\sigma_3(\partial_0 \Phi)U_R\right]=\Tr\left[\Phi^\dag\sigma_3(\partial_0 \Phi)\right]
\end{align}
and hence the Lagrangian is only invariant under left transformations for which $[\sigma_3,U_L]=0$, i.e. $U_L=\sigma_3$ (up to a normalization constant). 
Since there is only one symmetry generator (i.e. $\sigma_3$), the left algebra is reduced to $\rm SU(2)_L\to U(1)_L$. 
On the other hand, for right transformations we see (using cyclic property of the trace) that the full $\rm SU(2)_R$ remains. 
Hence the presence of the magnetic field $h\neq0$ acts to {\it explicitly} break the global symmetry: $\rm SU(2)_L\times SU(2)_R\to U(1)_L\times SU(2)_R$.\\

\noindent $\bullet$ When Bose condensation occurs, i.e. $\mu^2 h^2\geq\Delta_0^2$, the symmetry is spontaneously broken down to \mbox{$ \rm U(1)_L\times SU(2)_R\to U(1)$}. 
Finally, simple Goldstone counting would say that there are \mbox{[$\rm U(1)\times SU(2)/U(1)]=1+3-1=3$} Goldstone modes. 
However, Ref.~\onlinecite{scammell2015} shows that only two Goldstone modes arise---one linear and one quadratic. 
The presence of a quadratic Goldstone mode leads to different counting rules and we refer the reader to the original work \cite{nielson1976}. \\

\noindent $\bullet$ Including the gauge field $a_\mu$ gauges out a global U(1) symmetry and reduces the number of Goldstone modes. 
\cref{table:symmetry} summarizes the symmetry-breaking pattern and the number of Goldstone modes. 

\begin{table}[h]
\caption{Global Symmetry and Goldstone Modes}
\centering
\begin{tabular}{|c|c c |c c | c c|} %{|c|c|c|c|c|c|c|}
\hline\hline
\multirow{2}{*}{} & \multicolumn{2}{c|}{Symmetric Phase} & %
    \multicolumn{2}{c|}{ \  Condensate Phase \  } & \multicolumn{2}{c|}{Condensate + Gauge Field}\\
\cline{2-7}
 & $h=0$ & $h\neq0$ &\ \  $h=0$  \ \  & $h\neq0$ & \ \ \ \  $h=0$  \ \ \ & $h\neq0$ \\
\hline
 \ \ \ \ Global Symmetry \ & $\rm SU(2)\times SU(2)$ \   & \  $\rm U(1)\times SU(2)$ & SU(2) & U(1) & U(1) & -- \\

 \# of Goldstone Modes \ &0 & 0 & 3 & 2 & 2 & 1 \\
\hline
\end{tabular}
\label{table:symmetry}
\end{table}

%=================================================
\subsection{Properties of the spinon condensate and gauge field}

We now study the spinon condensate including the gauge field, i.e. the Lagrangian ${\cal L}[z,a_\mu]$ [\cref{Lza}].
The spinon condensate occurs by tuning $\mu^2 h^2> \Delta_0^2$, and we denote the corresponding non-zero expectation $\braket{z}\equiv z_0$. 
We choose the following parametrization 
\begin{align}
z_0=\begin{pmatrix}\cos\frac{\theta}{2}\\e^{i\phi}\sin\frac{\theta}{2}\end{pmatrix}\rho, \ \ \ \ \rho\in\mathbb{R}. 
\end{align}
In terms of these variables, energy density is particularly simple: 
\begin{align}
{\cal E}_0&=\Delta_0^2\rho^2+\frac{1}{2}\alpha \rho^4-\left[e^2\braket{a_0}^2\rho^2+2e \mu h \braket{a_0} \rho^2\cos\theta+ \mu^2 h^2\rho^2\right]+ e^2\braket{{\bm a}^2}\rho^2.
\end{align}
Here we use $a_\mu=(a_0,{\bm a})$. 
We can also find the classical expectation values of the gauge field from $\delta{\cal L}/\delta \braket{a_\mu}=0$
\begin{align}
\braket{a_\mu}&=-\frac{\mu h}{e}\cos\theta \delta_{\mu,0}.
\end{align}
Substituting this back into the energy density, we find
\begin{subequations}
\begin{align}
{\cal E}_0=&\left(\Delta_0^2- \mu^2 h^2\sin^2\theta\right)\rho^2+\frac{1}{2}\alpha \rho^4 , \\
\rho^2=&\frac{ \mu^2 h^2\sin^2\theta-\Delta_0^2}{\alpha}. 
\end{align}
\end{subequations}
From here we see that the spinon condensate takes the preferred direction $\theta=\pi/2$---i.e. zero density of the gauge field, $\braket{a_0}=0$.

%=================================================
\subsection{Spinon and gauge field fluctuations within the condensate \label{fluctuations}}

In this section we write the Lagrangian in terms of real field variables. 
We include fluctuations of the spinon condensate via the following parametrization 
\begin{align}
z&=e^{i \pi_s\sigma_s/\rho}\begin{pmatrix}1\\1 \end{pmatrix}\frac{\rho+H}{\sqrt{2}}\approx\begin{pmatrix} \rho+H+ i\pi_1 +\pi_2 +i\pi_3  \\ \rho+H+i\pi_1 - \pi_2 -i\pi_3 \end{pmatrix}\frac{1}{\sqrt{2}},
\end{align}
such that $\pi_s, s=1,2,3$ are the phase fluctuations (related to Goldstone modes), $H$ is the amplitude fluctuation, and $\sigma_s$ are Pauli matrices. 
All fields are real. 

First we consider the potential, which depends only on the $H$-mode. 
Using $|z|^2=(\rho+H)^2$ and $\rho^2=( \mu^2 h^2 -\Delta_0^2)/\alpha$ we get
\begin{align}
{\cal L}_{Potential}&=-(\Delta_0^2- \mu^2 h^2)|z|^2 -\frac{1}{2}\alpha|z|^4\Rightarrow -2\alpha \rho^2H^2 - \frac{1}{2}\alpha \left(4\rho H^3+ H^4\right),
\end{align}
where the ``$\Rightarrow$'' sign is used because irrelevant linear-in-$H$ terms are excluded (they are removed by the equations of motion). 
Next we consider the second order derivatives and rewrite in the real field variables:
\begin{align}
{\cal L}_{Quad}&=\left|D_\mu z\right|^2\approx\left|\frac{i}{\sqrt{2}}\begin{pmatrix} \partial_\mu \pi_1 + \partial_\mu \pi_3 - a_\mu \rho \\
\partial_\mu \pi_1 - \partial_\mu \pi_3 - a_\mu \rho \end{pmatrix}+ \frac{1}{\sqrt{2}}\begin{pmatrix} \partial_\mu \pi_2 + \partial_\mu H \\
-\partial_\mu \pi_2 + \partial_\mu H \end{pmatrix}\right|^2,\\
\notag&=(\tilde{a}_0 \rho)^2-(\tilde{{\bm a}}\rho)^2+(\partial_\mu \pi_2)^2+(\partial_\mu \pi_3)^2+(\partial_\mu H)^2.
\end{align}
Above $\partial_\mu^2=\partial_0^2-c^2\bar{\nabla}^2$.
Now we see that the gauge choice, 
\begin{align}
\label{gaugechoice}
\tilde{a}_\mu&=\partial_\mu \pi_1/\rho- a_\mu,
\end{align}
acts to remove $\pi_1$ kinetic energy. 
Next we rewrite the first order derivatives in the real field variables, and find
\begin{align}
{\cal L}_{Lin}&=\mu h z^\dag\sigma_3 D_0 z - \mu h (D_0 z)^\dag\sigma_3z= 4\mu h \pi_3\partial_0 H - 4\mu h \tilde{a}_0\rho \pi_2.
\end{align}
Here we have invoked the gauge choice \cref{gaugechoice}. 
Finally we obtain the full Lagrangian: 
\begin{align}
\label{L_real_gauge}
\notag&{\cal L}={\cal L}_{Quad} + {\cal L}_{Lin} + {\cal L}_{Potential} + {\cal L}_{Maxwell},\\
&{\cal L} =(\partial_\mu \pi_2)^2+(\partial_\mu \pi_3)^2+(\partial_\mu H)^2 + 4\mu h \pi_3\partial_0 H - 4\mu h\tilde{a}_0\rho \pi_2-2\alpha \rho^2H^2 - \frac{1}{2}\alpha \left(4\rho H^3+ H^4\right)+\tilde{a}_\mu^2 \rho^2 - \frac{1}{4}\tilde{f}_{\mu\nu}\tilde{f}^{\mu\nu}.
\end{align}
We see that the $\{\tilde{a}_\mu,\pi_2\}$ and $\{\pi_3,H\}$ sectors are decoupled from each other. 
{\it Comment 1}: Here the gauge charge is set to unity $e=1$. 
At the end of the calculation, factors of $e$ will be reinstated. 
{\it Comment 2}: Including $e$ we find that all fields have dimension $[a]=[H]=[\rho]=[\pi_i]=1/2$, and the charge itself has dimension $[e]=1/2$. 
Meanwhile, the interaction constant has dimension $[\alpha]=1$.

%=================================================
\subsection{Equations of motion and dispersions}
%=================================================
We now present the spectra of all modes, which provide insight into the physical origin of each of the real field fluctuations $\{\pi_1, \pi_2, \pi_3, H\}$.

\subsubsection{Higgs/Goldstone $\{H,\pi_3\}$ Sector}

The dispersions of the $\pi_3, H$ modes are 
\begin{align}
\omega_{1}&=\sqrt{3 \mu^2 h^2-\Delta_0^2+c^2k^2-\sqrt{(3 \mu^2 h^2-\Delta_0^2)^2+4 \mu^2 h^2c^2k^2}}\to\sqrt{\frac{ \mu^2 h^2-\Delta_0^2}{3 \mu^2 h^2-\Delta_0^2}}ck \ \ \ \text{(at $k\to0$)},\\
\omega_2&=\sqrt{3 \mu^2 h^2-\Delta_0^2+c^2k^2+\sqrt{(3 \mu^2 h^2-\Delta_0^2)^2+4 \mu^2 h^2c^2k^2}}\to\sqrt{6 \mu^2 h^2-2\Delta_0^2} \ \ \ \ \ \text{(at $k=0$)}.
\end{align}
To obtain these results, we work at tree-level, i.e. we exclude the higher than quadratic terms in the potential ${\cal L}_{\rm Potential}$ of \cref{L_real_gauge}.

%=================================================
\subsubsection{Goldstone/precession $\{\pi_1,\pi_2\}$ sector}

Looking first at $\{\pi_1,\pi_2\}$ in the absence of a gauge field:
\begin{align}
{\cal L}[\pi_1,\pi_2]&=(\partial_\mu \pi_1)^2+(\partial_\mu \pi_2)^2 + 2\mu h (\pi_1\partial_0\pi_2 - \pi_2\partial_0\pi_1). 
\end{align}
This sector gives the quadratic Goldstone mode and one of the Higgs-type modes (a precession mode): 
\begin{align}
\omega_k&=\sqrt{c^2k^2+ \mu^2 h^2}\pm \mu h.
\end{align}
Now we consider the gauge field and explicitly insert the gauge charge, $e$. 
First we note that in the disordered phase, gauge field only admits one mode, with dispersion $\omega= c k$. 
Due to the condensate, the gauge field acquires a ``mass'' term in the Lagrangian $\tilde{a}_\mu^2 e^2 \rho^2$, i.e. the Meisner effect for the emergent gauge field. 
In addition to the gauge field becoming massive, it also admits another mode.  
To proceed, we work in the Coulomb gauge $\bar{\nabla}\cdot{\bm a}=0$, and therefore we do not continue with the Goldstone absorption choice $\tilde{a}_\mu=\partial_\mu \pi_1/\rho- a_\mu$. 
Explicitly the Lagrangian is
\begin{align}
{\cal L}[\pi_1,\pi_2]=& (\partial_\mu \pi_1)^2+(\partial_\mu \pi_2)^2 + 2 \mu h (\pi_1\partial_0\pi_2 - \pi_2\partial_0\pi_1 + \pi_2a_0 \rho)+a_\mu^2 \rho^2 + \frac{1}{2}(\bar\nabla a_0)^2 + \frac{1}{2}(\partial_0 {\bm a})^2 - \frac{1}{2}(\nabla_i a_j)^2.
\end{align}
We then Fourier transform and rewrite in matrix form:
\begin{align}
\cal L&=\begin{pmatrix} \pi_1 \\ \pi_2 \\ a_0 \end{pmatrix}^T \left(
\begin{array}{ccc}
 \omega ^2-c^2k^2 & 2 i \mu h \omega  & i e \rho  \omega  \\
 -2 i \mu h \omega  & \omega ^2-c^2k^2 & 2 e \mu h \rho  \\
 -i e \rho \omega  & 2 e \mu h \rho & \frac{1}{2}c^2k^2+e^2\rho^2 \\
\end{array}
\right)\begin{pmatrix}\pi_1 \\ \pi_2 \\ a_0 \end{pmatrix} +\frac{1}{2} {\bm a}^T(
2e^2\rho^2 + c^2k^2-\omega ^2){\bm a}.
\end{align}

It is now straightforward to diagonalize the Lagrangian and obtain the dispersions. 
From the ${\bm a}$ sector of the Lagrangian, we obtain the gapped gauge field:
\begin{align}
\omega_5&=\sqrt{c^2k^2+2e^2\rho^2}\to\sqrt{2}e\rho \ \ \ \ \ \text{(at $k\to0$)}.
\end{align}
From the $\{\pi_1,\pi_2, a_0\}$ sector, we find that the precession and gapped Goldstone modes become
\begin{align}
\omega_{3,4}&=\sqrt{\mp\sqrt{\left(e^2\rho^2-2 \mu^2 h^2\right)^2+4 \mu^2 h^2 c^2k^2 }+2 \mu^2 h^2+e^2\rho^2+c^2k^2}.
\end{align}
Which have gaps: 
\begin{subequations}
\begin{align}
\label{sqconfine1}
\Delta_3& =\begin{cases}
  \sqrt{2}e\rho, & e^2\rho^2<2 \mu^2 h^2,\\    
2 \mu h, & e^2\rho^2>2 \mu^2 h^2.
\end{cases}\\
\label{sqconfine2}
\Delta_4& =\begin{cases}
  2 \mu h, & e^2\rho^2<2 \mu^2 h^2,\\    
\sqrt{2}e\rho, & e^2\rho^2>2 \mu^2 h^2, 
\end{cases}
\end{align}
\end{subequations}

%=================================================
\section{Magnon theory \label{supp:magnon}}

In this appendix we present the derivations for the magnon dispersions and mass renormalization in order to explain the predictions for the statistical energy of magnon BECs and gases in \cref{f:bec,f:sgas}, respectively. 
Original considerations can be found in Refs. \onlinecite{fisher1989,affleck1991}, and more extensive discussions in Ref.~\onlinecite{hsthesis}. 
The quantum phase transition between ordered and disordered phases is described by an effective field theory with the Lagrangian
\begin{align}   
\label{Cont_Lagrangian}
{\cal L}&=\frac{1}{2}(\partial_{t}{\vec{\varphi}}-\vec{\varphi}\times\mu\vec{h})^2-\frac{1}{2}(\vec{\nabla}{\vec{\varphi}})^2-\frac{1}{2}m^2_0{\vec{\varphi}}^{\ 2}-\frac{1}{4}\alpha_0\vec{\varphi}^{\ 4}.
\end{align}
Here, the vector field $\vec{\varphi}$ describes staggered magnetization, $\vec{h}$ is an external applied field, the magnetic moment is $\mu=1/2$, and we set $g\mu_B = c = 1$.

%------------------------------------------------
\subsection{Disordered magnon gas}
In the disordered phase, we can use the Euler-Lagrange equation and \cref{Cont_Lagrangian} to find the dispersion:
\begin{align}
\label{Cont_Zeeman}
\omega_\sigma&=\sqrt{k^2+m_{\Lambda,\sigma}^2}+\sigma \mu h,
\end{align}
with $\sigma=\pm$. 
In this phase, we consider the renormalization of the mass term. 
Let us denote by ${\cal V}$ the part of the Lagrangian [\cref{Cont_Lagrangian}] independent of derivatives. 
We then use Wick decoupling of the interaction term $\frac{1}{4}\alpha_0\vec{\varphi}^{\ 4}$ in the single-loop approximation to find
\begin{align}
m_\perp^2(T,h)-\mu^2 h^2=\notag\frac{\partial^2{\cal V}}{\partial\varphi_x^2}&=m_0^2-\mu^2 h^2+3\alpha_0\braket{\varphi_x^2} + \alpha_0\braket{\varphi_y^2} +\alpha_0\braket{\varphi_z^2}\\
m_\perp^2(T,h)-\mu^2 h^2=\notag\frac{\partial^2{\cal V}}{\partial\varphi_y^2}&=m_0^2-\mu^2 h^2+\alpha_0\braket{\varphi_x^2} + 3\alpha_0\braket{\varphi_y^2} +\alpha_0\braket{\varphi_z^2}\\
\label{Cont_curveRG}
m_z^2(T,h)=\frac{\partial^2{\cal V}}{\partial\varphi_z^2}&=m_0^2+\alpha_0\braket{\varphi_x^2} + \alpha_0\braket{\varphi_y^2} +3\alpha_0\braket{\varphi_z^2}
\end{align}
where $\braket{\varphi_x^2}$ is the loop integral over the Green's function of field $\varphi_x$. 
We absorb the zero-temperature divergence and consider the thermal contributions given by 
\begin{align}
\braket{\varphi_x^2}=\braket{\varphi_y^2}&=\int \frac{d^2l}{(2\pi c)^2}\frac{1}{\omega_0}\left(2n(\omega^+_l) + 2n(\omega^-_l) + n(\omega^0_l\right),\\
\braket{\varphi_z^2}&=\int \frac{d^2l}{(2\pi c)^2}\frac{1}{\omega_0}\left(n(\omega^+_l) + n(\omega^-_l) +3 n(\omega^0_l\right).
\end{align}
Again, $\omega^\pm_l=\omega_0(l)\pm \mu h$ and $\omega_0(l)=\sqrt{l^2+\Delta^2}$. 
Such integrals can be performed exactly, but we will leave them in this form. 

%------------------------------------------------
\subsection{Ordered magnon BEC}
In the magnon BEC phase $(h>h_c(T))$ the vector field is written $\vec\varphi=(\varphi_c+\sigma, \pi, z)$, where $\varphi_c$ is the order parameter field and fields $\sigma$ and $\pi$ correspond to hybridizations of the true Higgs and Goldstone modes. 
The field $z$ directly corresponds to the precession mode. 
The physical (diagonal) modes of the system have dispersions
\begin{subequations}
\begin{align}
\label{dispSIGMA}
\omega_{H}&=\sqrt{k^2+3\mu^2 h^2-m_{\Lambda,H}^2+\sqrt{4\mu^2 h^2k^2+(3\mu^2 h^2-m_{\Lambda,H}^2)^2}}\ ,\\
\label{dispPI}
\omega_{G}&=\sqrt{k^2+3\mu^2 h^2-m_{\Lambda,H}^2-\sqrt{4\mu^2 h^2k^2+(3\mu^2 h^2-m_{\Lambda,H}^2)^2}}\ ,\\
\label{dispZ}
\omega_{z}&=\sqrt{k^2+m^2_{\Lambda,z}}\ .
\end{align} 
\end{subequations}
Here the superscripts $\left\{ H, G, z\right\}$ designate the Higgs, Goldstone, and precession modes, respectively. In the limit $k\to0$, we obtain 
\begin{subequations}
\begin{align}
\label{limSIGMA}
\omega_{H}&\to\sqrt{6\mu^2 h^2-2m_{\Lambda,H}^2}\ ,\\
\label{limPI}
\omega_{G}&\to\sqrt{\frac{\mu^2 h^2 - m_{\Lambda,H}^2}{3\mu^2 h^2 - m_{\Lambda,H}^2}}c k\ ,\\
\label{limZ}
\omega_{z}&\to m_{\Lambda,z}\ .
\end{align} 
\end{subequations}

%=================================================
\end{widetext}
%=================================================

%=================================================
%include bibliography
\bibliography{bibstuff}

%apsrev4-2.bst 2019-01-14 (MD) hand-edited version of apsrev4-1.bst
%Control: key (0)
%Control: author (72) initials jnrlst
%Control: editor formatted (1) identically to author
%Control: production of article title (-1) disabled
%Control: page (0) single
%Control: year (1) truncated
%Control: production of eprint (0) enabled
\begin{thebibliography}{47}%
\makeatletter
\providecommand \@ifxundefined [1]{%
 \@ifx{#1\undefined}
}%
\providecommand \@ifnum [1]{%
 \ifnum #1\expandafter \@firstoftwo
 \else \expandafter \@secondoftwo
 \fi
}%
\providecommand \@ifx [1]{%
 \ifx #1\expandafter \@firstoftwo
 \else \expandafter \@secondoftwo
 \fi
}%
\providecommand \natexlab [1]{#1}%
\providecommand \enquote  [1]{``#1''}%
\providecommand \bibnamefont  [1]{#1}%
\providecommand \bibfnamefont [1]{#1}%
\providecommand \citenamefont [1]{#1}%
\providecommand \href@noop [0]{\@secondoftwo}%
\providecommand \href [0]{\begingroup \@sanitize@url \@href}%
\providecommand \@href[1]{\@@startlink{#1}\@@href}%
\providecommand \@@href[1]{\endgroup#1\@@endlink}%
\providecommand \@sanitize@url [0]{\catcode `\\12\catcode `\$12\catcode
  `\&12\catcode `\#12\catcode `\^12\catcode `\_12\catcode `\%12\relax}%
\providecommand \@@startlink[1]{}%
\providecommand \@@endlink[0]{}%
\providecommand \url  [0]{\begingroup\@sanitize@url \@url }%
\providecommand \@url [1]{\endgroup\@href {#1}{\urlprefix }}%
\providecommand \urlprefix  [0]{URL }%
\providecommand \Eprint [0]{\href }%
\providecommand \doibase [0]{https://doi.org/}%
\providecommand \selectlanguage [0]{\@gobble}%
\providecommand \bibinfo  [0]{\@secondoftwo}%
\providecommand \bibfield  [0]{\@secondoftwo}%
\providecommand \translation [1]{[#1]}%
\providecommand \BibitemOpen [0]{}%
\providecommand \bibitemStop [0]{}%
\providecommand \bibitemNoStop [0]{.\EOS\space}%
\providecommand \EOS [0]{\spacefactor3000\relax}%
\providecommand \BibitemShut  [1]{\csname bibitem#1\endcsname}%
\let\auto@bib@innerbib\@empty
%</preamble>
\bibitem [{\citenamefont {Sandvik}(2007)}]{sandvik2007}%
  \BibitemOpen
  \bibfield  {author} {\bibinfo {author} {\bibfnamefont {A.~W.}\ \bibnamefont
  {Sandvik}},\ }\href {https://doi.org/10.1103/PhysRevLett.98.227202}
  {\bibfield  {journal} {\bibinfo  {journal} {Phys. Rev. Lett.}\ }\textbf
  {\bibinfo {volume} {98}},\ \bibinfo {pages} {227202} (\bibinfo {year}
  {2007})}\BibitemShut {NoStop}%
\bibitem [{\citenamefont {Melko}\ and\ \citenamefont {Kaul}(2008)}]{melko2008}%
  \BibitemOpen
  \bibfield  {author} {\bibinfo {author} {\bibfnamefont {R.~G.}\ \bibnamefont
  {Melko}}\ and\ \bibinfo {author} {\bibfnamefont {R.~K.}\ \bibnamefont
  {Kaul}},\ }\href {https://doi.org/10.1103/PhysRevLett.100.017203} {\bibfield
  {journal} {\bibinfo  {journal} {Phys. Rev. Lett.}\ }\textbf {\bibinfo
  {volume} {100}},\ \bibinfo {pages} {017203} (\bibinfo {year}
  {2008})}\BibitemShut {NoStop}%
\bibitem [{\citenamefont {Lou}\ \emph {et~al.}(2009)\citenamefont {Lou},
  \citenamefont {Sandvik},\ and\ \citenamefont {Kawashima}}]{lou2009}%
  \BibitemOpen
  \bibfield  {author} {\bibinfo {author} {\bibfnamefont {J.}~\bibnamefont
  {Lou}}, \bibinfo {author} {\bibfnamefont {A.~W.}\ \bibnamefont {Sandvik}},\
  and\ \bibinfo {author} {\bibfnamefont {N.}~\bibnamefont {Kawashima}},\ }\href
  {https://doi.org/10.1103/PhysRevB.80.180414} {\bibfield  {journal} {\bibinfo
  {journal} {Phys. Rev. B}\ }\textbf {\bibinfo {volume} {80}},\ \bibinfo
  {pages} {180414(R)} (\bibinfo {year} {2009})}\BibitemShut {NoStop}%
\bibitem [{\citenamefont {Jin}\ and\ \citenamefont {Sandvik}(2013)}]{jin2013}%
  \BibitemOpen
  \bibfield  {author} {\bibinfo {author} {\bibfnamefont {S.}~\bibnamefont
  {Jin}}\ and\ \bibinfo {author} {\bibfnamefont {A.~W.}\ \bibnamefont
  {Sandvik}},\ }\href {https://doi.org/10.1103/PhysRevB.87.180404} {\bibfield
  {journal} {\bibinfo  {journal} {Phys. Rev. B}\ }\textbf {\bibinfo {volume}
  {87}},\ \bibinfo {pages} {180404(R)} (\bibinfo {year} {2013})}\BibitemShut
  {NoStop}%
\bibitem [{\citenamefont {Chen}\ \emph {et~al.}(2013)\citenamefont {Chen},
  \citenamefont {Huang}, \citenamefont {Deng}, \citenamefont {Kuklov},
  \citenamefont {Prokof'ev},\ and\ \citenamefont {Svistunov}}]{chen2013}%
  \BibitemOpen
  \bibfield  {author} {\bibinfo {author} {\bibfnamefont {K.}~\bibnamefont
  {Chen}}, \bibinfo {author} {\bibfnamefont {Y.}~\bibnamefont {Huang}},
  \bibinfo {author} {\bibfnamefont {Y.}~\bibnamefont {Deng}}, \bibinfo {author}
  {\bibfnamefont {A.~B.}\ \bibnamefont {Kuklov}}, \bibinfo {author}
  {\bibfnamefont {N.~V.}\ \bibnamefont {Prokof'ev}},\ and\ \bibinfo {author}
  {\bibfnamefont {B.~V.}\ \bibnamefont {Svistunov}},\ }\href
  {https://doi.org/10.1103/PhysRevLett.110.185701} {\bibfield  {journal}
  {\bibinfo  {journal} {Phys. Rev. Lett.}\ }\textbf {\bibinfo {volume} {110}},\
  \bibinfo {pages} {185701} (\bibinfo {year} {2013})}\BibitemShut {NoStop}%
\bibitem [{\citenamefont {Pujari}\ \emph {et~al.}(2013)\citenamefont {Pujari},
  \citenamefont {Damle},\ and\ \citenamefont {Alet}}]{pujari2013}%
  \BibitemOpen
  \bibfield  {author} {\bibinfo {author} {\bibfnamefont {S.}~\bibnamefont
  {Pujari}}, \bibinfo {author} {\bibfnamefont {K.}~\bibnamefont {Damle}},\ and\
  \bibinfo {author} {\bibfnamefont {F.}~\bibnamefont {Alet}},\ }\href
  {https://doi.org/10.1103/PhysRevLett.111.087203} {\bibfield  {journal}
  {\bibinfo  {journal} {Phys. Rev. Lett.}\ }\textbf {\bibinfo {volume} {111}},\
  \bibinfo {pages} {087203} (\bibinfo {year} {2013})}\BibitemShut {NoStop}%
\bibitem [{\citenamefont {Harada}\ \emph {et~al.}(2013)\citenamefont {Harada},
  \citenamefont {Suzuki}, \citenamefont {Okubo}, \citenamefont {Matsuo},
  \citenamefont {Lou}, \citenamefont {Watanabe}, \citenamefont {Todo},\ and\
  \citenamefont {Kawashima}}]{harada2013}%
  \BibitemOpen
  \bibfield  {author} {\bibinfo {author} {\bibfnamefont {K.}~\bibnamefont
  {Harada}}, \bibinfo {author} {\bibfnamefont {T.}~\bibnamefont {Suzuki}},
  \bibinfo {author} {\bibfnamefont {T.}~\bibnamefont {Okubo}}, \bibinfo
  {author} {\bibfnamefont {H.}~\bibnamefont {Matsuo}}, \bibinfo {author}
  {\bibfnamefont {J.}~\bibnamefont {Lou}}, \bibinfo {author} {\bibfnamefont
  {H.}~\bibnamefont {Watanabe}}, \bibinfo {author} {\bibfnamefont
  {S.}~\bibnamefont {Todo}},\ and\ \bibinfo {author} {\bibfnamefont
  {N.}~\bibnamefont {Kawashima}},\ }\href
  {https://doi.org/10.1103/PhysRevB.88.220408} {\bibfield  {journal} {\bibinfo
  {journal} {Phys. Rev. B}\ }\textbf {\bibinfo {volume} {88}},\ \bibinfo
  {pages} {220408(R)} (\bibinfo {year} {2013})}\BibitemShut {NoStop}%
\bibitem [{\citenamefont {Shao}\ \emph {et~al.}(2016)\citenamefont {Shao},
  \citenamefont {Guo},\ and\ \citenamefont {Sandvik}}]{shao2016}%
  \BibitemOpen
  \bibfield  {author} {\bibinfo {author} {\bibfnamefont {H.}~\bibnamefont
  {Shao}}, \bibinfo {author} {\bibfnamefont {W.}~\bibnamefont {Guo}},\ and\
  \bibinfo {author} {\bibfnamefont {A.~W.}\ \bibnamefont {Sandvik}},\ }\href
  {https://doi.org/10.1126/science.aad5007} {\bibfield  {journal} {\bibinfo
  {journal} {Science}\ }\textbf {\bibinfo {volume} {352}},\ \bibinfo {pages}
  {213} (\bibinfo {year} {2016})}\BibitemShut {NoStop}%
\bibitem [{\citenamefont {Senthil}\ \emph
  {et~al.}(2004{\natexlab{a}})\citenamefont {Senthil}, \citenamefont {Balents},
  \citenamefont {Sachdev}, \citenamefont {Vishwanath},\ and\ \citenamefont
  {Fisher}}]{senthil2004}%
  \BibitemOpen
  \bibfield  {author} {\bibinfo {author} {\bibfnamefont {T.}~\bibnamefont
  {Senthil}}, \bibinfo {author} {\bibfnamefont {L.}~\bibnamefont {Balents}},
  \bibinfo {author} {\bibfnamefont {S.}~\bibnamefont {Sachdev}}, \bibinfo
  {author} {\bibfnamefont {A.}~\bibnamefont {Vishwanath}},\ and\ \bibinfo
  {author} {\bibfnamefont {M.~P.~A.}\ \bibnamefont {Fisher}},\ }\href
  {https://doi.org/10.1103/PhysRevB.70.144407} {\bibfield  {journal} {\bibinfo
  {journal} {Phys. Rev. B}\ }\textbf {\bibinfo {volume} {70}},\ \bibinfo
  {pages} {144407} (\bibinfo {year} {2004}{\natexlab{a}})}\BibitemShut
  {NoStop}%
\bibitem [{\citenamefont {Senthil}\ \emph
  {et~al.}(2004{\natexlab{b}})\citenamefont {Senthil}, \citenamefont
  {Vishwanath}, \citenamefont {Balents}, \citenamefont {Sachdev},\ and\
  \citenamefont {Fisher}}]{senthil}%
  \BibitemOpen
  \bibfield  {author} {\bibinfo {author} {\bibfnamefont {T.}~\bibnamefont
  {Senthil}}, \bibinfo {author} {\bibfnamefont {A.}~\bibnamefont {Vishwanath}},
  \bibinfo {author} {\bibfnamefont {L.}~\bibnamefont {Balents}}, \bibinfo
  {author} {\bibfnamefont {S.}~\bibnamefont {Sachdev}},\ and\ \bibinfo {author}
  {\bibfnamefont {M.~P.~A.}\ \bibnamefont {Fisher}},\ }\href
  {https://doi.org/10.1126/science.1091806} {\bibfield  {journal} {\bibinfo
  {journal} {Science}\ }\textbf {\bibinfo {volume} {303}},\ \bibinfo {pages}
  {1490} (\bibinfo {year} {2004}{\natexlab{b}})}\BibitemShut {NoStop}%
\bibitem [{\citenamefont {Ma}\ \emph {et~al.}(2018)\citenamefont {Ma},
  \citenamefont {Sun}, \citenamefont {You}, \citenamefont {Xu}, \citenamefont
  {Vishwanath}, \citenamefont {Sandvik},\ and\ \citenamefont {Meng}}]{ma2018}%
  \BibitemOpen
  \bibfield  {author} {\bibinfo {author} {\bibfnamefont {N.}~\bibnamefont
  {Ma}}, \bibinfo {author} {\bibfnamefont {G.-Y.}\ \bibnamefont {Sun}},
  \bibinfo {author} {\bibfnamefont {Y.-Z.}\ \bibnamefont {You}}, \bibinfo
  {author} {\bibfnamefont {C.}~\bibnamefont {Xu}}, \bibinfo {author}
  {\bibfnamefont {A.}~\bibnamefont {Vishwanath}}, \bibinfo {author}
  {\bibfnamefont {A.~W.}\ \bibnamefont {Sandvik}},\ and\ \bibinfo {author}
  {\bibfnamefont {Z.~Y.}\ \bibnamefont {Meng}},\ }\href
  {https://doi.org/10.1103/PhysRevB.98.174421} {\bibfield  {journal} {\bibinfo
  {journal} {Phys. Rev. B}\ }\textbf {\bibinfo {volume} {98}},\ \bibinfo
  {pages} {174421} (\bibinfo {year} {2018})}\BibitemShut {NoStop}%
\bibitem [{\citenamefont {Scammell}\ and\ \citenamefont
  {Sushkov}(2015)}]{scammell2015}%
  \BibitemOpen
  \bibfield  {author} {\bibinfo {author} {\bibfnamefont {H.~D.}\ \bibnamefont
  {Scammell}}\ and\ \bibinfo {author} {\bibfnamefont {O.~P.}\ \bibnamefont
  {Sushkov}},\ }\href {https://doi.org/10.1103/PhysRevLett.114.055702}
  {\bibfield  {journal} {\bibinfo  {journal} {Phys. Rev. Lett.}\ }\textbf
  {\bibinfo {volume} {114}},\ \bibinfo {pages} {055702} (\bibinfo {year}
  {2015})}\BibitemShut {NoStop}%
\bibitem [{\citenamefont {Fisher}(1989)}]{fisher1989}%
  \BibitemOpen
  \bibfield  {author} {\bibinfo {author} {\bibfnamefont {D.~S.}\ \bibnamefont
  {Fisher}},\ }\href {https://doi.org/10.1103/PhysRevB.39.11783} {\bibfield
  {journal} {\bibinfo  {journal} {Phys. Rev. B}\ }\textbf {\bibinfo {volume}
  {39}},\ \bibinfo {pages} {11783} (\bibinfo {year} {1989})}\BibitemShut
  {NoStop}%
\bibitem [{\citenamefont {Affleck}(1991)}]{affleck1991}%
  \BibitemOpen
  \bibfield  {author} {\bibinfo {author} {\bibfnamefont {I.}~\bibnamefont
  {Affleck}},\ }\href {https://doi.org/10.1103/PhysRevB.43.3215} {\bibfield
  {journal} {\bibinfo  {journal} {Phys. Rev. B}\ }\textbf {\bibinfo {volume}
  {43}},\ \bibinfo {pages} {3215} (\bibinfo {year} {1991})}\BibitemShut
  {NoStop}%
\bibitem [{\citenamefont {Majumdar}\ and\ \citenamefont
  {Ghosh}(1969{\natexlab{a}})}]{majumdar1}%
  \BibitemOpen
  \bibfield  {author} {\bibinfo {author} {\bibfnamefont {C.~K.}\ \bibnamefont
  {Majumdar}}\ and\ \bibinfo {author} {\bibfnamefont {D.~K.}\ \bibnamefont
  {Ghosh}},\ }\href {https://doi.org/10.1063/1.1664978} {\bibfield  {journal}
  {\bibinfo  {journal} {J. Math. Phys.}\ }\textbf {\bibinfo {volume} {10}},\
  \bibinfo {pages} {1388} (\bibinfo {year} {1969}{\natexlab{a}})}\BibitemShut
  {NoStop}%
\bibitem [{\citenamefont {Majumdar}\ and\ \citenamefont
  {Ghosh}(1969{\natexlab{b}})}]{majumdar2}%
  \BibitemOpen
  \bibfield  {author} {\bibinfo {author} {\bibfnamefont {C.~K.}\ \bibnamefont
  {Majumdar}}\ and\ \bibinfo {author} {\bibfnamefont {D.~K.}\ \bibnamefont
  {Ghosh}},\ }\href {https://doi.org/10.1063/1.1664979} {\bibfield  {journal}
  {\bibinfo  {journal} {J. Math. Phys.}\ }\textbf {\bibinfo {volume} {10}},\
  \bibinfo {pages} {1399} (\bibinfo {year} {1969}{\natexlab{b}})}\BibitemShut
  {NoStop}%
\bibitem [{\citenamefont {Haldane}(1982)}]{haldane1982}%
  \BibitemOpen
  \bibfield  {author} {\bibinfo {author} {\bibfnamefont {F.~D.~M.}\
  \bibnamefont {Haldane}},\ }\href {https://doi.org/10.1103/PhysRevB.25.4925}
  {\bibfield  {journal} {\bibinfo  {journal} {Phys. Rev. B}\ }\textbf {\bibinfo
  {volume} {25}},\ \bibinfo {pages} {4925} (\bibinfo {year}
  {1982})}\BibitemShut {NoStop}%
\bibitem [{\citenamefont {Haldane}(1988)}]{haldane1988}%
  \BibitemOpen
  \bibfield  {author} {\bibinfo {author} {\bibfnamefont {F.~D.~M.}\
  \bibnamefont {Haldane}},\ }\href
  {https://doi.org/10.1103/PhysRevLett.61.1029} {\bibfield  {journal} {\bibinfo
   {journal} {Phys. Rev. Lett.}\ }\textbf {\bibinfo {volume} {61}},\ \bibinfo
  {pages} {1029} (\bibinfo {year} {1988})}\BibitemShut {NoStop}%
\bibitem [{\citenamefont {Read}\ and\ \citenamefont
  {Sachdev}(1989)}]{read1989}%
  \BibitemOpen
  \bibfield  {author} {\bibinfo {author} {\bibfnamefont {N.}~\bibnamefont
  {Read}}\ and\ \bibinfo {author} {\bibfnamefont {S.}~\bibnamefont {Sachdev}},\
  }\href {https://doi.org/10.1103/PhysRevLett.62.1694} {\bibfield  {journal}
  {\bibinfo  {journal} {Phys. Rev. Lett.}\ }\textbf {\bibinfo {volume} {62}},\
  \bibinfo {pages} {1694} (\bibinfo {year} {1989})}\BibitemShut {NoStop}%
\bibitem [{\citenamefont {Sachdev}(2008)}]{sachdev2008}%
  \BibitemOpen
  \bibfield  {author} {\bibinfo {author} {\bibfnamefont {S.}~\bibnamefont
  {Sachdev}},\ }\href {https://doi.org/10.1038/nphys894} {\bibfield  {journal}
  {\bibinfo  {journal} {Nat. Phys.}\ }\textbf {\bibinfo {volume} {4}},\
  \bibinfo {pages} {173} (\bibinfo {year} {2008})}\BibitemShut {NoStop}%
\bibitem [{\citenamefont {Dagotto}\ and\ \citenamefont
  {Moreo}(1989)}]{dagotto1989}%
  \BibitemOpen
  \bibfield  {author} {\bibinfo {author} {\bibfnamefont {E.}~\bibnamefont
  {Dagotto}}\ and\ \bibinfo {author} {\bibfnamefont {A.}~\bibnamefont
  {Moreo}},\ }\href {https://doi.org/10.1103/PhysRevLett.63.2148} {\bibfield
  {journal} {\bibinfo  {journal} {Phys. Rev. Lett.}\ }\textbf {\bibinfo
  {volume} {63}},\ \bibinfo {pages} {2148} (\bibinfo {year}
  {1989})}\BibitemShut {NoStop}%
\bibitem [{\citenamefont {{H.J. Schulz}}\ \emph {et~al.}(1996)\citenamefont
  {{H.J. Schulz}}, \citenamefont {{T.A.L. Ziman}},\ and\ \citenamefont {{D.
  Poilblanc}}}]{schultz1996}%
  \BibitemOpen
  \bibfield  {author} {\bibinfo {author} {\bibnamefont {{H.J. Schulz}}},
  \bibinfo {author} {\bibnamefont {{T.A.L. Ziman}}},\ and\ \bibinfo {author}
  {\bibnamefont {{D. Poilblanc}}},\ }\href
  {https://doi.org/10.1051/jp1:1996236} {\bibfield  {journal} {\bibinfo
  {journal} {J. Phys. I France}\ }\textbf {\bibinfo {volume} {6}},\ \bibinfo
  {pages} {675} (\bibinfo {year} {1996})}\BibitemShut {NoStop}%
\bibitem [{\citenamefont {Suwa}\ \emph {et~al.}(2016)\citenamefont {Suwa},
  \citenamefont {Sen},\ and\ \citenamefont {Sandvik}}]{suwa2016}%
  \BibitemOpen
  \bibfield  {author} {\bibinfo {author} {\bibfnamefont {H.}~\bibnamefont
  {Suwa}}, \bibinfo {author} {\bibfnamefont {A.}~\bibnamefont {Sen}},\ and\
  \bibinfo {author} {\bibfnamefont {A.~W.}\ \bibnamefont {Sandvik}},\ }\href
  {https://doi.org/10.1103/PhysRevB.94.144416} {\bibfield  {journal} {\bibinfo
  {journal} {Phys. Rev. B}\ }\textbf {\bibinfo {volume} {94}},\ \bibinfo
  {pages} {144416} (\bibinfo {year} {2016})}\BibitemShut {NoStop}%
\bibitem [{\citenamefont {Wang}\ \emph {et~al.}(2017)\citenamefont {Wang},
  \citenamefont {Nahum}, \citenamefont {Metlitski}, \citenamefont {Xu},\ and\
  \citenamefont {Senthil}}]{wang2017}%
  \BibitemOpen
  \bibfield  {author} {\bibinfo {author} {\bibfnamefont {C.}~\bibnamefont
  {Wang}}, \bibinfo {author} {\bibfnamefont {A.}~\bibnamefont {Nahum}},
  \bibinfo {author} {\bibfnamefont {M.~A.}\ \bibnamefont {Metlitski}}, \bibinfo
  {author} {\bibfnamefont {C.}~\bibnamefont {Xu}},\ and\ \bibinfo {author}
  {\bibfnamefont {T.}~\bibnamefont {Senthil}},\ }\href
  {https://doi.org/10.1103/PhysRevX.7.031051} {\bibfield  {journal} {\bibinfo
  {journal} {Phys. Rev. X}\ }\textbf {\bibinfo {volume} {7}},\ \bibinfo {pages}
  {031051} (\bibinfo {year} {2017})}\BibitemShut {NoStop}%
\bibitem [{\citenamefont {Sandvik}\ \emph {et~al.}(2011)\citenamefont
  {Sandvik}, \citenamefont {Kotov},\ and\ \citenamefont
  {Sushkov}}]{sandvik2011}%
  \BibitemOpen
  \bibfield  {author} {\bibinfo {author} {\bibfnamefont {A.~W.}\ \bibnamefont
  {Sandvik}}, \bibinfo {author} {\bibfnamefont {V.~N.}\ \bibnamefont {Kotov}},\
  and\ \bibinfo {author} {\bibfnamefont {O.~P.}\ \bibnamefont {Sushkov}},\
  }\href {https://doi.org/10.1103/PhysRevLett.106.207203} {\bibfield  {journal}
  {\bibinfo  {journal} {Phys. Rev. Lett.}\ }\textbf {\bibinfo {volume} {106}},\
  \bibinfo {pages} {207203} (\bibinfo {year} {2011})}\BibitemShut {NoStop}%
\bibitem [{\citenamefont {Sandvik}(2010)}]{sandvik2011computational}%
  \BibitemOpen
  \bibfield  {author} {\bibinfo {author} {\bibfnamefont {A.~W.}\ \bibnamefont
  {Sandvik}},\ }\href {https://doi.org/10.1063/1.3518900} {\bibfield  {journal}
  {\bibinfo  {journal} {AIP Conf. Proc.}\ }\textbf {\bibinfo {volume} {1297}},\
  \bibinfo {pages} {135} (\bibinfo {year} {2010})}\BibitemShut {NoStop}%
\bibitem [{\citenamefont {Sylju\aa{}sen}\ and\ \citenamefont
  {Sandvik}(2002)}]{sandvik_dl}%
  \BibitemOpen
  \bibfield  {author} {\bibinfo {author} {\bibfnamefont {O.~F.}\ \bibnamefont
  {Sylju\aa{}sen}}\ and\ \bibinfo {author} {\bibfnamefont {A.~W.}\ \bibnamefont
  {Sandvik}},\ }\href {https://doi.org/10.1103/PhysRevE.66.046701} {\bibfield
  {journal} {\bibinfo  {journal} {Phys. Rev. E}\ }\textbf {\bibinfo {volume}
  {66}},\ \bibinfo {pages} {046701} (\bibinfo {year} {2002})}\BibitemShut
  {NoStop}%
\bibitem [{\citenamefont {Sandvik}(2002)}]{sandvik2002}%
  \BibitemOpen
  \bibfield  {author} {\bibinfo {author} {\bibfnamefont {A.~W.}\ \bibnamefont
  {Sandvik}},\ }\href {https://doi.org/10.1103/PhysRevB.66.024418} {\bibfield
  {journal} {\bibinfo  {journal} {Phys. Rev. B}\ }\textbf {\bibinfo {volume}
  {66}},\ \bibinfo {pages} {024418} (\bibinfo {year} {2002})}\BibitemShut
  {NoStop}%
\bibitem [{\citenamefont {Iaizzi}\ and\ \citenamefont
  {Sandvik}(2015)}]{iaizzi2015}%
  \BibitemOpen
  \bibfield  {author} {\bibinfo {author} {\bibfnamefont {A.}~\bibnamefont
  {Iaizzi}}\ and\ \bibinfo {author} {\bibfnamefont {A.~W.}\ \bibnamefont
  {Sandvik}},\ }\href {http://stacks.iop.org/1742-6596/640/i=1/a=012043}
  {\bibfield  {journal} {\bibinfo  {journal} {J. Phys. Conf. Ser.}\ }\textbf
  {\bibinfo {volume} {640}},\ \bibinfo {pages} {012043} (\bibinfo {year}
  {2015})}\BibitemShut {NoStop}%
\bibitem [{\citenamefont {Iaizzi}\ \emph {et~al.}(2017)\citenamefont {Iaizzi},
  \citenamefont {Damle},\ and\ \citenamefont {Sandvik}}]{iaizzi2017}%
  \BibitemOpen
  \bibfield  {author} {\bibinfo {author} {\bibfnamefont {A.}~\bibnamefont
  {Iaizzi}}, \bibinfo {author} {\bibfnamefont {K.}~\bibnamefont {Damle}},\ and\
  \bibinfo {author} {\bibfnamefont {A.~W.}\ \bibnamefont {Sandvik}},\ }\href
  {https://doi.org/10.1103/PhysRevB.95.174436} {\bibfield  {journal} {\bibinfo
  {journal} {Phys. Rev. B}\ }\textbf {\bibinfo {volume} {95}},\ \bibinfo
  {pages} {174436} (\bibinfo {year} {2017})}\BibitemShut {NoStop}%
\bibitem [{\citenamefont {Iaizzi}\ \emph {et~al.}(2018)\citenamefont {Iaizzi},
  \citenamefont {Damle},\ and\ \citenamefont {Sandvik}}]{iaizzi2018metamag}%
  \BibitemOpen
  \bibfield  {author} {\bibinfo {author} {\bibfnamefont {A.}~\bibnamefont
  {Iaizzi}}, \bibinfo {author} {\bibfnamefont {K.}~\bibnamefont {Damle}},\ and\
  \bibinfo {author} {\bibfnamefont {A.~W.}\ \bibnamefont {Sandvik}},\ }\href
  {https://doi.org/10.1103/PhysRevB.98.064405} {\bibfield  {journal} {\bibinfo
  {journal} {Phys. Rev. B}\ }\textbf {\bibinfo {volume} {98}},\ \bibinfo
  {pages} {064405} (\bibinfo {year} {2018})}\BibitemShut {NoStop}%
\bibitem [{\citenamefont {Iaizzi}(2018)}]{mythesis}%
  \BibitemOpen
  \bibfield  {author} {\bibinfo {author} {\bibfnamefont {A.}~\bibnamefont
  {Iaizzi}},\ }\href {https://doi.org/10.1007/978-3-030-01803-0} {\emph
  {\bibinfo {title} {Magnetic Field Effects in Low-Dimensional Quantum
  Magnets}}},\ Springer Theses\ (\bibinfo  {publisher} {Springer},\ \bibinfo
  {year} {2018})\BibitemShut {NoStop}%
\bibitem [{\citenamefont {Kosterlitz}\ and\ \citenamefont
  {Thouless}(1972)}]{kosterlitz1972}%
  \BibitemOpen
  \bibfield  {author} {\bibinfo {author} {\bibfnamefont {J.~M.}\ \bibnamefont
  {Kosterlitz}}\ and\ \bibinfo {author} {\bibfnamefont {D.~J.}\ \bibnamefont
  {Thouless}},\ }\href {http://stacks.iop.org/0022-3719/5/i=11/a=002}
  {\bibfield  {journal} {\bibinfo  {journal} {J. Phys. C}\ }\textbf {\bibinfo
  {volume} {5}},\ \bibinfo {pages} {L124} (\bibinfo {year} {1972})}\BibitemShut
  {NoStop}%
\bibitem [{\citenamefont {Landau}\ and\ \citenamefont
  {Binder}(1981)}]{landau1981}%
  \BibitemOpen
  \bibfield  {author} {\bibinfo {author} {\bibfnamefont {D.~P.}\ \bibnamefont
  {Landau}}\ and\ \bibinfo {author} {\bibfnamefont {K.}~\bibnamefont
  {Binder}},\ }\href {https://doi.org/10.1103/PhysRevB.24.1391} {\bibfield
  {journal} {\bibinfo  {journal} {Phys. Rev. B}\ }\textbf {\bibinfo {volume}
  {24}},\ \bibinfo {pages} {1391} (\bibinfo {year} {1981})}\BibitemShut
  {NoStop}%
\bibitem [{\citenamefont {Pires}(1994)}]{pires1994}%
  \BibitemOpen
  \bibfield  {author} {\bibinfo {author} {\bibfnamefont {A.~S.~T.}\
  \bibnamefont {Pires}},\ }\href {https://doi.org/10.1103/PhysRevB.50.9592}
  {\bibfield  {journal} {\bibinfo  {journal} {Phys. Rev. B}\ }\textbf {\bibinfo
  {volume} {50}},\ \bibinfo {pages} {9592} (\bibinfo {year}
  {1994})}\BibitemShut {NoStop}%
\bibitem [{\citenamefont {Cuccoli}\ \emph {et~al.}(2003)\citenamefont
  {Cuccoli}, \citenamefont {Roscilde}, \citenamefont {Vaia},\ and\
  \citenamefont {Verrucchi}}]{cuccoli2003}%
  \BibitemOpen
  \bibfield  {author} {\bibinfo {author} {\bibfnamefont {A.}~\bibnamefont
  {Cuccoli}}, \bibinfo {author} {\bibfnamefont {T.}~\bibnamefont {Roscilde}},
  \bibinfo {author} {\bibfnamefont {R.}~\bibnamefont {Vaia}},\ and\ \bibinfo
  {author} {\bibfnamefont {P.}~\bibnamefont {Verrucchi}},\ }\href
  {https://doi.org/10.1103/PhysRevB.68.060402} {\bibfield  {journal} {\bibinfo
  {journal} {Phys. Rev. B}\ }\textbf {\bibinfo {volume} {68}},\ \bibinfo
  {pages} {060402(R)} (\bibinfo {year} {2003})}\BibitemShut {NoStop}%
\bibitem [{\citenamefont {Cuccoli}\ \emph {et~al.}(2004)\citenamefont
  {Cuccoli}, \citenamefont {Roscilde}, \citenamefont {Vaia},\ and\
  \citenamefont {Verrucchi}}]{cuccoli2004}%
  \BibitemOpen
  \bibfield  {author} {\bibinfo {author} {\bibfnamefont {A.}~\bibnamefont
  {Cuccoli}}, \bibinfo {author} {\bibfnamefont {T.}~\bibnamefont {Roscilde}},
  \bibinfo {author} {\bibfnamefont {R.}~\bibnamefont {Vaia}},\ and\ \bibinfo
  {author} {\bibfnamefont {P.}~\bibnamefont {Verrucchi}},\ }\href
  {https://doi.org/10.1016/j.jmmm.2003.12.605} {\bibfield  {journal} {\bibinfo
  {journal} {J. Magn. Magn. Mater.}\ }\textbf {\bibinfo {volume} {272–276,
  Part 2}},\ \bibinfo {pages} {884 } (\bibinfo {year} {2004})}\BibitemShut
  {NoStop}%
\bibitem [{\citenamefont {Baranov\'{a}}\ \emph {et~al.}(2016)\citenamefont
  {Baranov\'{a}}, \citenamefont {Orend\'{a}\v{c}ov\'{a}}, \citenamefont
  {\v{C}i\v{z}m\'{a}r}, \citenamefont {Tarasenko}, \citenamefont {Tk\'a\v{c}},
  \citenamefont {Orend\'a\v{c}},\ and\ \citenamefont {Feher}}]{baranova2016}%
  \BibitemOpen
  \bibfield  {author} {\bibinfo {author} {\bibfnamefont {L.}~\bibnamefont
  {Baranov\'{a}}}, \bibinfo {author} {\bibfnamefont {A.}~\bibnamefont
  {Orend\'{a}\v{c}ov\'{a}}}, \bibinfo {author} {\bibfnamefont {E.}~\bibnamefont
  {\v{C}i\v{z}m\'{a}r}}, \bibinfo {author} {\bibfnamefont {R.}~\bibnamefont
  {Tarasenko}}, \bibinfo {author} {\bibfnamefont {V.}~\bibnamefont
  {Tk\'a\v{c}}}, \bibinfo {author} {\bibfnamefont {M.}~\bibnamefont
  {Orend\'a\v{c}}},\ and\ \bibinfo {author} {\bibfnamefont {A.}~\bibnamefont
  {Feher}},\ }\href {https://doi.org/10.1016/j.jmmm.2015.12.025} {\bibfield
  {journal} {\bibinfo  {journal} {J. Magn. Magn. Mater.}\ }\textbf {\bibinfo
  {volume} {404}},\ \bibinfo {pages} {53 } (\bibinfo {year}
  {2016})}\BibitemShut {NoStop}%
\bibitem [{\citenamefont {Qin}\ \emph {et~al.}(2017)\citenamefont {Qin},
  \citenamefont {He}, \citenamefont {You}, \citenamefont {Lu}, \citenamefont
  {Sen}, \citenamefont {Sandvik}, \citenamefont {Xu},\ and\ \citenamefont
  {Meng}}]{qin2017}%
  \BibitemOpen
  \bibfield  {author} {\bibinfo {author} {\bibfnamefont {Y.~Q.}\ \bibnamefont
  {Qin}}, \bibinfo {author} {\bibfnamefont {Y.-Y.}\ \bibnamefont {He}},
  \bibinfo {author} {\bibfnamefont {Y.-Z.}\ \bibnamefont {You}}, \bibinfo
  {author} {\bibfnamefont {Z.-Y.}\ \bibnamefont {Lu}}, \bibinfo {author}
  {\bibfnamefont {A.}~\bibnamefont {Sen}}, \bibinfo {author} {\bibfnamefont
  {A.~W.}\ \bibnamefont {Sandvik}}, \bibinfo {author} {\bibfnamefont
  {C.}~\bibnamefont {Xu}},\ and\ \bibinfo {author} {\bibfnamefont {Z.~Y.}\
  \bibnamefont {Meng}},\ }\href {https://doi.org/10.1103/PhysRevX.7.031052}
  {\bibfield  {journal} {\bibinfo  {journal} {Phys. Rev. X}\ }\textbf {\bibinfo
  {volume} {7}},\ \bibinfo {pages} {031052} (\bibinfo {year}
  {2017})}\BibitemShut {NoStop}%
\bibitem [{\citenamefont {Ma}\ \emph {et~al.}(2019)\citenamefont {Ma},
  \citenamefont {You},\ and\ \citenamefont {Meng}}]{ma2019}%
  \BibitemOpen
  \bibfield  {author} {\bibinfo {author} {\bibfnamefont {N.}~\bibnamefont
  {Ma}}, \bibinfo {author} {\bibfnamefont {Y.-Z.}\ \bibnamefont {You}},\ and\
  \bibinfo {author} {\bibfnamefont {Z.~Y.}\ \bibnamefont {Meng}},\ }\href
  {https://doi.org/10.1103/PhysRevLett.122.175701} {\bibfield  {journal}
  {\bibinfo  {journal} {Phys. Rev. Lett.}\ }\textbf {\bibinfo {volume} {122}},\
  \bibinfo {pages} {175701} (\bibinfo {year} {2019})}\BibitemShut {NoStop}%
\bibitem [{\citenamefont {Nelson}\ and\ \citenamefont
  {Kosterlitz}(1977)}]{nelson1977}%
  \BibitemOpen
  \bibfield  {author} {\bibinfo {author} {\bibfnamefont {D.~R.}\ \bibnamefont
  {Nelson}}\ and\ \bibinfo {author} {\bibfnamefont {J.~M.}\ \bibnamefont
  {Kosterlitz}},\ }\href {https://doi.org/10.1103/PhysRevLett.39.1201}
  {\bibfield  {journal} {\bibinfo  {journal} {Phys. Rev. Lett.}\ }\textbf
  {\bibinfo {volume} {39}},\ \bibinfo {pages} {1201} (\bibinfo {year}
  {1977})}\BibitemShut {NoStop}%
\bibitem [{\citenamefont {Hsieh}\ \emph {et~al.}(2013)\citenamefont {Hsieh},
  \citenamefont {Kao},\ and\ \citenamefont {Sandvik}}]{hsieh2013}%
  \BibitemOpen
  \bibfield  {author} {\bibinfo {author} {\bibfnamefont {Y.-D.}\ \bibnamefont
  {Hsieh}}, \bibinfo {author} {\bibfnamefont {Y.-J.}\ \bibnamefont {Kao}},\
  and\ \bibinfo {author} {\bibfnamefont {A.~W.}\ \bibnamefont {Sandvik}},\
  }\href {http://stacks.iop.org/1742-5468/2013/i=09/a=P09001} {\bibfield
  {journal} {\bibinfo  {journal} {J. Stat. Mech.}\ }\textbf {\bibinfo {volume}
  {2013}},\ \bibinfo {pages} {P09001} (\bibinfo {year} {2013})}\BibitemShut
  {NoStop}%
\bibitem [{\citenamefont {{Song}}\ \emph {et~al.}(2019)\citenamefont {{Song}},
  \citenamefont {{Wang}}, \citenamefont {{Vishwanath}},\ and\ \citenamefont
  {{He}}}]{song2018}%
  \BibitemOpen
  \bibfield  {author} {\bibinfo {author} {\bibfnamefont {X.-Y.}\ \bibnamefont
  {{Song}}}, \bibinfo {author} {\bibfnamefont {C.}~\bibnamefont {{Wang}}},
  \bibinfo {author} {\bibfnamefont {A.}~\bibnamefont {{Vishwanath}}},\ and\
  \bibinfo {author} {\bibfnamefont {Y.-C.}\ \bibnamefont {{He}}},\ }\href
  {https://doi.org/10.1038/s41467-019-11727-3} {\bibfield  {journal} {\bibinfo
  {journal} {Nat. Comm.}\ }\textbf {\bibinfo {volume} {10}},\ \bibinfo {pages}
  {4254} (\bibinfo {year} {2019})}\BibitemShut {NoStop}%
\bibitem [{\citenamefont {Srednicki}(2007)}]{Srednicki}%
  \BibitemOpen
  \bibfield  {author} {\bibinfo {author} {\bibfnamefont {M.}~\bibnamefont
  {Srednicki}},\ }\href {https://books.google.com.tw/books?id=5OepxIG42B4C}
  {\emph {\bibinfo {title} {Quantum Field Theory}}}\ (\bibinfo  {publisher}
  {Cambridge University Press},\ \bibinfo {year} {2007})\BibitemShut {NoStop}%
\bibitem [{\citenamefont {Blaizot}\ and\ \citenamefont
  {Iancu}(1996)}]{blaizot1996}%
  \BibitemOpen
  \bibfield  {author} {\bibinfo {author} {\bibfnamefont {J.-P.}\ \bibnamefont
  {Blaizot}}\ and\ \bibinfo {author} {\bibfnamefont {E.}~\bibnamefont
  {Iancu}},\ }\href {https://doi.org/10.1016/0550-3213(95)00547-1} {\bibfield
  {journal} {\bibinfo  {journal} {Nuclear Physics B}\ }\textbf {\bibinfo
  {volume} {459}},\ \bibinfo {pages} {559 } (\bibinfo {year}
  {1996})}\BibitemShut {NoStop}%
\bibitem [{\citenamefont {Nielsen}\ and\ \citenamefont
  {Chadha}(1976)}]{nielson1976}%
  \BibitemOpen
  \bibfield  {author} {\bibinfo {author} {\bibfnamefont {H.}~\bibnamefont
  {Nielsen}}\ and\ \bibinfo {author} {\bibfnamefont {S.}~\bibnamefont
  {Chadha}},\ }\href {https://doi.org/10.1016/0550-3213(76)90025-0} {\bibfield
  {journal} {\bibinfo  {journal} {Nuclear Physics B}\ }\textbf {\bibinfo
  {volume} {105}},\ \bibinfo {pages} {445 } (\bibinfo {year}
  {1976})}\BibitemShut {NoStop}%
\bibitem [{\citenamefont {Scammell}(2018)}]{hsthesis}%
  \BibitemOpen
  \bibfield  {author} {\bibinfo {author} {\bibfnamefont {H.}~\bibnamefont
  {Scammell}},\ }\href {https://doi.org/10.1007/978-3-319-97532-0} {\emph
  {\bibinfo {title} {Interplay of Quantum and Statistical Fluctuations in
  Critical Quantum Matter}}},\ Springer Theses\ (\bibinfo  {publisher}
  {Springer},\ \bibinfo {year} {2018})\BibitemShut {NoStop}%
\end{thebibliography}%

\end{document}